\documentclass[a4paper,10pt]{article}
\pdfoutput=1

\usepackage{jheppub}
\usepackage[utf8]{inputenc}
\usepackage{amsmath, amsfonts}
\usepackage{graphicx}
\usepackage[all]{xy}

\usepackage{xcolor,colortbl}

\newcommand{\tr}{\mathrm{tr}}
\newcommand{\bO}{\mathcal{O}}
\newcommand{\A}{\mathcal{A}}
\newcommand{\C}{\mathcal{C}}
\newcommand{\eps}{\varepsilon}

\title{Magnetic field induced lattice ground states from holography}

\author[a,b]{Yan-Yan Bu,}
\author[a]{Johanna Erdmenger,}
\author[a]{Jonathan P. Shock,}
\author[a]{Migael Strydom}

\affiliation[a]{Max-Planck-Institut f\"{u}r Physik (Werner-Heisenberg-Institut), \\
F\"{o}rhringer Ring 6, 80805 M\"{u}nchen, Germany.
}
\affiliation[b]{State Key Laboratory of Theoretical Physics,
Institute of Theoretical Physics, Chinese Academy of Science, Beijing 100190, People's Republic of China
}

\emailAdd{ (yybu,jke,jonshock,mstrydom)@mppmu.mpg.de}

\keywords{Gauge-gravity correspondence, Holography and condensed matter
physics(AdS/CMT) }

\preprint{MPP-2012-144}

\abstract{
We study  the holographic field theory dual of a probe $SU(2)$ Yang-Mills field
in a background $(4+1)$-dimensional asymptotically Anti-de Sitter space.
We find a new ground state when a magnetic component of the gauge field is
larger than a critical value. The ground state forms a
triangular Abrikosov lattice in the spatial directions perpendicular to the
magnetic field. The lattice is composed of superconducting vortices induced by
the
condensation of a charged vector operator. We perform this calculation both at
finite temperature and at zero temperature with a hard wall cutoff dual to a
confining gauge theory. The study of this state may
be of relevance to both holographic condensed matter models as well as to
heavy ion physics. The results shown here provide support for the
proposal that such a ground state may be found in the QCD
vacuum when a large magnetic field is present.
}

\begin{document}

\maketitle

\section{Introduction}
The study of black hole instabilities is an important research topic that has
led to very interesting results.
In particular, within gauge/gravity duality, the study of
Anti-de Sitter black hole solutions and their stability properties is
important for understanding thermal states on the gauge theory side.
Last year in~\cite{Ammon:2011je}, some of the authors of the present
paper studied an $SU(2)$ Einstein-Yang-Mills model at finite
temperature in asymptotically AdS space. They found that when a
magnetic component of the gauge field reaches a critical value in
units of the temperature, the system becomes unstable. Though the
critical value of the magnetic field for onset of the instability was
calculated, the new ground state of the system was not known. In the
current work we calculate a ground state solution, using a
perturbative analysis similar to the one performed by Abrikosov
in~\cite{Abrikosov:1956sx} for type II superconductors. In agreement
with the work of Abrikosov we find that the ground state is a
triangular lattice. There have been many attempts recently to model
lattices holographically with the goal of providing more realistic
models for condensed matter systems
\cite{Horowitz:2012ky,Flauger:2010tv}, and this novel procedure for
generating a lattice dynamically adds to these developments. Moreover,
our holographic model provides support for recent QCD studies of
$\rho$ meson condensation from a strong magnetic field
\cite{Chernodub:2010qx,Chernodub:2010zw,Chernodub:2011gs}. The effect
described here is similar to the Nielsen-Olsen solution for gluon
condensation \cite{Nielsen:1978rm} and to magnetically catalysed
W~boson condensation
\cite{Ambjorn:1988tm,Ambjorn:1989bd,Ambjorn:1989sz}.

Although this is the first holographic calculation to explicitly
uncover an Abrikosov lattice in 3+1 dimensions, it is not the first to
examine spatially inhomogeneous phases of strongly coupled field
theories. In \cite{Horowitz:2012ky}, the authors studied the holographic construction of an
Einstein-Maxwell-scalar theory at finite temperature and
density. They looked at the gauge theory optical conductivity, which is the conductivity in the direction of an applied electric field. They broke the translational invariance explicitly by
imposing scalar field boundary conditions in the form of a lattice modulated in
one of the Minkowski spatial directions. A fully backreacted solution
was found which thus induces a spatially inhomogeneous black hole
solution. This leads to an extremely rich behaviour of the frequency dependent optical conductivity. At low frequencies there appears
a Drude peak. A Drude
peak is a broadening of the zero frequency delta peak in the conductivity. In real materials this is due to impurities and finite
temperature effects. The Drude peak is not present when translational
invariance is unbroken. The solution also exhibits a power law
behaviour at frequencies intermediate with respect to the temperature, and a constant value in the high
frequency regime. The power law behaviour is the same as that found
experimentally in cuprates, while the constant value at high
frequencies is expected from conformal invariance. The setup in this
context was a (2+1)-dimensional model where the lattice was periodic
in only one of the spatial dimensions. A more realistic lattice
structure would be highly desirable.

While the lattice in the approach of~\cite{Horowitz:2012ky} was
implemented in the boundary conditions, there are a number of other
mechanisms known that lead dynamically to ground states without
translational symmetry. One approach was pioneered in
\cite{Domokos:2007kt} by studying a Yang-Mills-Chern-Simons theory in
the gauge/gravity context. It was shown that the Chern-Simons term
can induce an instability which leads to a ground state with both
translational and rotational symmetry breaking. Such work was
continued in
\cite{Nakamura:2009tf,Chuang:2010ku,Bergman:2011rf,Bayona:2011ab,
Takeuchi:2011uk} where the spatially homogeneous phase was found to be
unstable in a variety of gravitational contexts in the presence of
Chern-Simons couplings. The perturbative analysis of quasinormal modes
that become tachyonic at finite momentum gives a relatively simple
computational tool for finding instabilities to ground states without
translational symmetry. These solutions are found to induce a helical
current \cite{Ooguri:2010kt,Donos:2012gg,Donos:2012wi}. Interestingly,
a Chern-Simons term is not always enough to induce such an
instability. It was shown by~\cite{Ammon:2011hz} that this type of
instability does not exist in the D3/D7 system.

Translational invariance can also be broken with a magnetic field or with magnetic monopoles. The former was first studied by Gauntlett et al. in~\cite{Donos:2011qt} and~\cite{Donos:2011bh} where the instability of the magnetically charged black hole
in a top-down framework was studied in detail. In the latter work, an infinite
family of solutions coming from $D=11$ supergravity was shown to exhibit a
magnetically catalysed instability. Such work is important as it proves that
these instabilities can also come from real string theory constructions.
The subject of magnetic monopoles in $(3+1)$-dimensional AdS space was studied
in~\cite{Bolognesi:2010nb} and~\cite{Sutcliffe:2011sr}. These magnetic monopoles
are solutions to the scalar field in a Yang-Mills-Higgs theory with gauge group
$SU(2)$.
In a certain limit where the monopole magnetic charge becomes large and a ``monopole wall'' is formed, it was shown in~\cite{Bolognesi:2010nb} that there is a W boson instability. In~\cite{Sutcliffe:2011sr} a hexagonal lattice ground state of these monopole walls was found numerically.
In~\cite{Allahbakhshi:2011nh} the holographic dual of a self-gravitating Julia-Zee Dyon was constructed, and it was shown to contain a vortex condensate.

There are some holographic models exhibiting a superconducting phase transition that results in a vortex lattice ground state. The first we mention involves an $s$-wave superconductor.
In~\cite{Maeda:2009vf} a type II superconductor was modelled using a $(3+1)$-dimensional gravitational setup. A type
II superconductor is one for which the external applied magnetic field
has two critical values. When the magnitude of the magnetic field
increases beyond the lower of the two critical values, the field starts to
penetrate the superconducting condensate. Some of the condensate
remains until the magnitude of the field is increased beyond the
upper critical value, at which point superconductivity is completely
destroyed. Just before the upper critical value is reached from
below, the ground state of the system is a triangular Abrikosov
lattice~\cite{Abrikosov:1956sx}. The authors of~\cite{Maeda:2009vf}
constructed a holographic superconductor modelling the behaviour of a type II superconductor near the upper critical value of the magnetic field and
found the Abrikosov lattice ground state explicitly.\footnote{
The model of~\cite{Maeda:2009vf} does not display the transition at the lower critical magnetic field value because the gauge field is not dynamical. See~\cite{Domenech:2010nf} for adding dynamical gauge fields to holographic superconductors. The transition at the upper critical value is present however because there the condensate is small so the backreaction is negligible.
}
In~\cite{Murray:2011gr} it was shown how to construct a similar vortex lattice solution in a model describing a $p$-wave superconductor. There the authors used a holographic model with an $SU(2)$ gauge field similar to the one described in the current paper.
Both of theses examples are different
from our model, however, because here we find a superconducting
Abrikosov lattice ground state that is induced by an $SU(2)$ magnetic
field, rather than being destroyed by it. Moreover, in contrast to
these models, we do not need a finite density. Our model is a cousin of holographic p-wave superconductors where the condensation is induced by a finite isospin density, holographically realised by a non-trivial temporal component of the $SU(2)$ gauge field (see \cite{Gubser:2008wv} and \cite{Ammon:2008fc,Ammon:2009fe} as well as the recent \cite{Chunlen:2012zy}). Here, in contrast, a spatial component of this gauge field has a non-trivial profile. Whereas in \cite{Ammon:2008fc,Ammon:2009fe}, a Meissner effect is shown to occur by which a magnetic field reduces the transition temperature, here it is again the magnetic field which induces condensation at zero density.

In addition to being interesting in the broader context of holographic
lattices, the model we discuss serves as supporting evidence for a
phenomenon first described by Chernodub et al. in~\cite{Chernodub:2010qx,Chernodub:2010zw}. There it was proposed that
the QCD$\times$QED vacuum may itself be susceptible to a
superconducting transition when a magnetic field of the order of the
QCD scale is present. Such extreme conditions are rare but they may be
present for a few femtoseconds during highly off-centre heavy ion
collisions. The discovery of this phase came about through the study
of an effective field theory description (the DSGS model proposed by
Djukanovic, Schindler, Gegelia and Scherer in
\cite{Djukanovic:2005ag}) of $\rho$ mesons interacting with a
magnetic field. A destabilisation of the vacuum was shown that would
clearly lead to the condensation of charged and neutral $\rho$ mesons.
This breaks the $U(1)$ gauge symmetry and leads to a superconductor
with the quark-antiquark pairs in the mesons acting as Cooper pairs.
The instability was also found using an extended Nambu--Jona-Lasinio
model with $SU(3)$ colour and $SU(2)$ flavour in
\cite{Chernodub:2011mc}. Lattice gauge theory studies were then
performed looking at QCD in strong magnetic fields and these
indicate the same instability. Moreover, using the DSGS model and
guided by the Ginzburg-Landau model of type II superconductors, a solution was found in which the
$\rho$ meson condensate forms an Abrikosov lattice made up of
superconducting vortices \cite{Chernodub:2011gs}. This may be relevant
experimentally. Evidence has mounted at both RHIC and the ALICE experiment at CERN that
strong magnetic fields may contribute to the physics of the strongly coupled
quark gluon plasma as charges are quickly accelerated during the
interaction period \cite{Skokov:2009qp,Bzdak:2011yy}. The importance
of these effects remains a contested topic because the time scales involved are small.
However, given that strong magnetic fields may be present, it is interesting
to ask if traces of this $\rho$ meson condensate could be detected.

In the current paper we find a possible ground state of the system in
\cite{Ammon:2011je}. As mentioned above, this system was shown to
be unstable under the imposition of a large $SU(2)$ magnetic
field.\footnote{It was shown in~\cite{Callebaut:2011ab} that the same
sort of instability occurs in the Sakai-Sugimoto model, but there the
ground state has also not been found.} We show that it has very
similar properties to the ground state of a type II superconductor
near the upper critical magnetic field as well as to the ground state in the
model of Chernodub et al. In other words, the ground state is a
triangular Abrikosov lattice. We take here a very simple model of a
strongly coupled finite temperature quantum field theory in
$(3+1)$-dimensions with a global $SU(2)$ symmetry. The dual gravity
theory is an $SU(2)$ Einstein-Yang-Mills theory in $(4+1)$-dimensions
with a magnetic component of the $SU(2)$ switched on. We work entirely
within the probe approximation, which means that the Yang-Mills
term is small compared to the Einstein-Hilbert term in the action.
We also fix the gauge in such a way that the gauge theory condensate is
transformed under a $U(1)$ subgroup of the global $SU(2)$ symmetry. There
appears to be a certain universality to the triangular lattice ground
state. Here we show that it forms in both the AdS Schwarzschild background (dual
to a finite temperature field theory) as well as the hard wall cutoff
model (dual to a confining field theory). It would be interesting to uncover exactly how universal
these results are.

The two holographic models that we study have several important differences
from QCD. In the finite temperature model there is no confinement or chiral symmetry
breaking and so there are no goldstone bosons (pions) present which
are the normal decay modes of the $\rho$ meson in QCD. The hard wall model has its conformal symmetry broken only by an IR boundary condition which sets a confinement scale. However, the phenomenology of these two models appears to be close enough to that of QCD to compare qualitatively with the models of Chernodub et al.

In section~\ref{sec:holset} we provide the details of the holographic
setup. There we also explain the strategy behind the perturbative expansion
of the $SU(2)$ gauge field near the critical magnetic field. Since we
follow the philosophy of Abrikosov's calculation of the ground state
in type II superconductors, which was done in the Ginzburg-Landau
model, in section~\ref{sec:solequ} we give a brief outline of this
approach and then follow it to solve perturbatively up to third order.
In section~\ref{sec:result} we discuss the numerical results and
analyse the free energy of the different lattice solutions, showing
that the triangular lattice has the lowest free energy of all the
Abrikosov solutions studied. It is important to note that we are not
able to show conclusively that we have found {\emph {the}} ground
state but we are able to find a state with lower free energy than the
translationally invariant state and that has lowest energy within a
large class of lattice solutions. In section~\ref{sec:conclu} we
conclude and give an outline of important future work.

\section{Holographic setup}\label{sec:holset}
\subsection{The finite temperature and hard wall backgrounds}
The system we study is an Einstein-Yang-Mills theory on the
(Poincar\'e patch of) an asymptotically AdS$_5$ geometry with an $SU(2)$ gauge
field. The action is
\begin{equation}
  S = \int d^5 x \sqrt{-g}~\left\{
    \frac{1}{16\pi G_N} \left(R + \frac{12}{L^2}\right)
    - \frac{1}{4 \hat{g}^2} \tr\left(F_{\mu \nu} F^{\mu\nu}\right)
  \right\}~,
  \label{eq:actioneym}
\end{equation}
where $\hat{g}$ is the Yang-Mills coupling, $G_N$ is the 5D gravitational
constant and $L$ is the AdS$_5$ radius. $R$ and $F$ are the Ricci
scalar and Yang-Mills field strength respectively.

We consider the probe approximation, where the Yang-Mills term is small
compared to the Einstein-Hilbert term, so that the backreaction of the gauge
fields on the geometry can be neglected. We thus choose a fixed 5-dimensional
background metric, given by
\begin{equation}
  ds^2
  =
  \frac{L^2}{u^2} \left(
    -f(u) dt^2 + dx^2 + dy^2 + dz^2 + \frac{du^2}{f(u)}
  \right)~,
  \label{eq:bfmetric}
\end{equation}
where the asymptotically AdS region is at $u\rightarrow 0$.
We study two different models. The first is a finite temperature model
where the background is AdS Schwarzschild, first proposed
in~\cite{Witten:1998qj}. In this case, $f(u)=1-\frac{u^4}{u_H^4}$,
where $u_H$ is the location of the planar black hole horizon. The Hawking
temperature of the black hole is $T=1/\pi u_H$. The second model is
the hard wall cutoff model, proposed in~\cite{Erlich:2005qh, DaRold:2005zs},
where
$f(u)=1$ and the geometry terminates at a radial distance $u_C$. This
model corresponds to a zero temperature theory ($u_H=\infty$), but it
still has a scale $u_C$ which corresponds to a confinement scale in
the gauge theory. The intrinsic scales in these theories allow us to form a dimensionless magnitude for the magnetic field. This will be the parameter that we tune in order to find the instability of the spatially invariant ground state. Without loss of generality
we can choose units where $u_H=1$ in the finite temperature theory and
$u_C=1$ in the confining theory. Factors of $u_H$ and $u_C$ can then be
restored through dimensional analysis. In the following the exact form
of the metric is not important until we come to solving the numerical
equations in the radial direction of AdS.
\subsection{The Yang-Mills action}
The relevant part of the action simplifies to
\begin{equation}
  S = - \frac{1}{4 \hat{g}^2} \int d^5 x \sqrt{-g}~
     \tr\left(F_{\mu \nu} F^{\mu\nu}\right)~,
  \label{eq:actionym}
\end{equation}
with the equations of motion
\begin{equation}
  \nabla^\mu F^a_{\mu \nu} + \epsilon^{abc} \A^{b\mu} F^c_{\mu \nu} = 0~.
  \label{eq:eomF}
\end{equation}
The $SU(2)$ gauge field is $\A = \A_\mu^a \tau^a dx^\mu$, for $a=1\dots 3$.
We use the convention where the Lie algebra basis is given by
$\tau^a=\frac{\sigma^a}{2i}$, with $\sigma^a$ the Pauli matrices, and the
structure constants $f^{abc}$ are defined by
$[\tau^a,\tau^b]= \epsilon^{abc} \tau^c$ so that $f^{abc} = \epsilon^{abc}$.
With
these definitions, the components of the field-strength tensor
$F=d\A + \A\wedge \A$ become
\begin{equation}
  F^a_{\mu \nu}
  =
  \partial_\mu \A^a_\nu - \partial_\nu \A^a_\mu
  +
  \epsilon^{abc} \A^b_\mu \A^c_\nu~.
\end{equation}

It will be important to understand how gauge transformations
affect the system. Under a gauge transformation $e^{i \Lambda(x^\mu)}$, $\A$
transforms as
\begin{equation}
  \A_\mu
  \rightarrow
  \A_\mu + \delta \A_\mu
  =
  e^{i \Lambda} \A_\mu e^{-i \Lambda}
  -
  i \partial_\mu e^{i\Lambda} e^{-i \Lambda}~.
\end{equation}
When $\Lambda(x^\mu)$ is an infinitesimal transformation,
this becomes
\begin{equation}
  \delta \A^a_\mu = \mathcal{D}_\mu \Lambda^a
  = \partial_\mu \Lambda^a + \epsilon^{abc} \A^b_\mu \Lambda^c~.
  \label{eq:gtinf}
\end{equation}
The gauge transformations give us the freedom to fix the gauge $\A^a_u=0$. We
 work in this gauge from now on.

In this paper we look at the effect of a strong (flavour-)magnetic field given by
$F^3_{xy}=B$, with all other components of $F^a_{\mu\nu}$ vanishing. As we will
see,
when $B$ becomes large\footnote{Since we have chosen the units where $u_H=1$ or
$u_C=1$, $B$ is a dimensionless quantity. Restoring the units, the statement is
that $B u_{H}^2=B/(\pi T)^2$ or $B u_C^2\sim B/\Lambda_{QCD}^2$ is large, or that
$B$ is large compared to the radial scale of the background.}, other components of
$F$ become non-zero dynamically. To get a consistent set of equations we therefore
consider a gauge field $\A$ of the form
\begin{equation} \A =
\sum_{a=1,2,3,\mu=x,y}\A^a_\mu(x,y,u)\tau^adx^\mu~. \label{eq:Achoice}
\end{equation}
It turns out that we can turn off the $t$ and $z$ dependence of the
gauge field and still have consistent equations. This simplifies the equations.
Turning off the $t$ dependence guarantees a static solution. Turning off
the $z$ dependence, where the $z$ direction is parallel to the magnetic field,
yields a lattice in the $x,y$-plane.

The action~\ref{eq:actionym} has an $SU(2)$ gauge freedom. Choosing the
solution $F^3_{xy}=B$, with all other components vanishing, breaks this
symmetry. Only $U(1)$ transformations of the form $\Lambda = \Lambda^3 \tau^3$
leave it invariant. For $B$ large enough, all the components
in~\ref{eq:Achoice} become nonzero due to the dynamics. We thus claim to have a
superconductor, because the $U(1)$ symmetry is broken dynamically. Note
however that it is technically a superfluid because the $U(1)$ gauge symmetry in
the bulk theory gets mapped to a global symmetry in the field theory. Taking
the linear combinations $\mathcal{E}^\pm_\mu= \A^1_\mu\pm i\A^2_\mu$ gives
fields that transform in the fundamental of the remaining gauge symmetry. It can
be checked from~\ref{eq:gtinf} that
$\mathcal{E}^\pm_\mu\rightarrow \mp i\Lambda^3\mathcal{E}^\pm_\mu$ whenever
$\Lambda = \Lambda^3 \tau^3$. Later on we work only with the fields
$\mathcal{E}^+$, which we rename to $\mathcal{E}$.

\subsection{Perturbative expansion of the gauge fields}

Substituting the ansatz~\ref{eq:Achoice} into equation~\ref{eq:eomF} yields nine
coupled partial differential equations in the variables $x$, $y$ and $u$. Of
these nine equations of motion, six are dynamical equations for each
field $\A^{1,2,3}_{x,y}$, and three equations are constraints. The constraint
equations arise from the equations of motion for the components $\A^{1,2,3}_u$,
which were chosen to be zero using gauge symmetry.

In solving the PDE's, we follow the strategy
of~\cite{Abrikosov:1956sx,AbrikosovBook}, which works as follows.
When the magnetic field $B$ is smaller than some
critical value $B_c$, the field configuration $\A^3_y = x B$,
$\A^3_x = 0$ and $\A^{1,2}_{x,y}=0$
solves the equations of motion. This is the
normal phase of the superconductor. As shown
in~\cite{Ammon:2011je}, the system
enters a new phase when the magnetic field is increased beyond some
critical value $B_c$. In this phase, the superconducting phase, the ground state
has a non-trivial profile
for all fields in the ansatz equation~\ref{eq:Achoice}. We look for this
configuration at some value of $B$ infinitesimally above $B_c$, where the
condensate is still small. This lets us do a perturbative expansion in a
small parameter $\eps\sim\frac{B-B_c}{B_c}$. For
notational convenience we leave this parameter $\eps$ explicit when studying the
expansion. However, it will be absorbed into the definition of the perturbative
corrections to the fields when we come to minimising the energy. We thus write
an ansatz for the expansion in the form
\begin{align}\label{eq:eqansatz}
  \A^3_y &= x B_c + \eps A^3_y + \eps^2 a^3_y + \dots, \\
  \A^a_\mu &= \eps A^a_\mu + \eps^2 a^a_\mu + \dots
  ~~~~~\mathrm{for~}(a,\mu) \ne (3,y)~,
\end{align}
and solve the equations order by order in $\eps$, as detailed in section~\ref{sec:solequ}.

\subsection{Gauge field boundary conditions}

The holographic dictionary relates field theory operators to gravity theory
fields through the relation
\begin{align}
  e^{-W_\mathrm{CFT}[\A^{(0)}]}
  =
  \langle e^{\int_{\partial AdS} \A^{(0)}_\mu J^\mu} \rangle
  =
  e^{-S_\mathrm{on-shell}}~.
  \label{eq:holdict}
\end{align}
The minus sign on the right-hand side is because we are in Euclidean space for
simplicity.
Here $\A^{(0)}$ is the value of the gauge field $\A$ at the AdS boundary. It
acts as a source in the boundary field theory. In our setup, the only source we
want in the field theory comes from the component $\A^3_y = x B$, producing the
magnetic field. For the other components in~\ref{eq:Achoice}, there should be
no explicit source because we want to model spontaneous symmetry breaking. The
spontaneous symmetry breaking results in a vev\footnote{We also need to take
holographic renormalisation into account to yield a finite on-shell action.},
\begin{align}
  \langle J^\mu \rangle
  =
  \left.
    \frac{\delta W_\mathrm{CFT}}{\delta \A^{(0)}_\mu}
  \right|_{\A^{(0)}_\mu=0}
  =
  \left.
    \frac{\delta S_\mathrm{on-shell}}{\delta \A^{(0)}_\mu}
  \right|_{\A^{(0)}_\mu=0}
  =
  \left.
  -\int d^4 x
  \frac{\partial \mathcal{L}}{\partial \left(\partial_u \A_\mu \right)}
  \right|_{u=0}
  \label{eq:vevJmu}
\end{align}
The second equality is a generalisation to the radial coordinate of one of the
steps in deriving the Hamilton-Jacobi equation. It relates the variation of
final value of a generalised coordinate with respect to the on-shell action and
the conjugate momentum at the final time.

It is interesting to note that the on-shell action can be written as
\begin{align}
  S_\mathrm{on-shell}
  =&
  -\frac{1}{2\hat{g}^2} \int_{\partial AdS} d^d x \sqrt{-\gamma}
    n_\mu A^a_\nu F^{a\mu\nu}
  + \frac{1}{4\hat{g}^2} \int_{AdS} d^{d+1} x \sqrt{-g}
  \epsilon^{abc} A^a_\mu A^b_\nu F^{c\mu\nu}~,
\end{align}
where we integrated by parts and substituted in the equations of motion. The
second term on the right-hand side, the bulk term, is not present in
non-interacting theories. In our case, however, it is present and nonzero even
after using ansatz~\ref{eq:Achoice}. This bulk term should seemingly influence
the calculation of the condensate when varying with respect to the boundary
value. It turns out that, due to the formula at the right of
equality~\ref{eq:vevJmu}, it makes no contribution.

Equations~\ref{eq:holdict} and~\ref{eq:vevJmu} imply that in an expansion of
the gauge fields near the AdS boundary, the leading term is the source and the
subleading term is proportional to the vev.
The field $\A_{x,y}^3$ has a boundary expansion given by
\begin{equation}
\left.\A_{x,y}^3\right|_{u \rightarrow 0}
=
s_{x,y}^{(3)}+v_{x,y}^{(3)} u^2+\dots~,
\end{equation}
where $s_{x,y}^{(3)}$ is the value of the source, which in this case is the
externally applied magnetic field potential. $v_{x,y}^{(3)}$ is proportional to
the vev corresponding to the magnetisation. We set the boundary
conditions so
that the applied magnetic field is not corrected by the higher order
perturbations
in $\eps$, whereas the magnetisation will obtain a non-zero value.

Similarly, the fields $A_{x,y}^{1,2}$ have a boundary expansion given by
\begin{equation}
\left.{\A_{x,y}^{1,2}}\right|_{u \rightarrow 0}
=
s_{x,y}^{(1,2)}+v_{x,y}^{(1,2)} u^2+\dots~,
\label{eq:A12exp}
\end{equation}
where $s^{(1,2)}_{x,y}$ corresponds to the source of the operator that will
condense to break the $U(1)$ symmetry. We adjust the boundary conditions in
such a way
that this vanishes. This means that the symmetry breaking is spontaneous.
$v_{x,y}^{(1,2)}$ is proportional to the vacuum expectation value of this
operator, which we
read off to find the resulting supercurrent in the superconducting phase.

Boundary conditions are also imposed on the fields in the IR. In the case of the
black hole background, we impose regularity at the horizon and in the case of
the
hard wall model we impose Neumann boundary conditions.

\subsection{The gauge theory ground state energy}\label{sec:energy}

In finding the ground state, it is important to be able to calculate the energy of
the field theory solution from the action. We would like to compare the
solutions
in the normal phase to those in the superconducting phase.
The energy $\mathcal{F}$ of the gauge theory solution is found by using the
holographic dictionary. In the case of the finite temperature solution, we are in
the canonical ensemble and we calculate the free energy, which is
$\mathcal{F}/T=-\ln \mathcal{Z}=- S_{cl}$ with our conventions. Here
$S_{cl}=-\frac{1}{4\hat{g}^2} \int d^5x \sqrt{-g} F^a_{\mu\nu}F^{a\mu\nu}$ is
the classical action. In the hard wall case, we are simply calculating the energy
of the field configuration, which is defined in terms of the classical action in the same way. Since we
are only
interested in whether the energy of a particular superconducting solution is lower
than that of the normal phase
solution, we can simply calculate the difference
$\Delta\mathcal{F}=\mathcal{F}_{s}-\mathcal{F}_{n}$ and thus do not need to
implement holographic renormalisation. Here $\mathcal{F}_s$ is
the energy of the superconducting phase, while
$\mathcal{F}_n$ is the normal phase energy with $\A^3_y=x B$
and all other components zero. We also need to take care
of the fact that $S_{cl}$ diverges when we perform the integral over the Minkowski
directions. This is easy to
fix by considering the energy density\footnote{We divide the free energy by $T$
in the finite temperature model to get a dimensionless $\Omega$. This means
that in both models, our total dimensionless energy is simply -$S_{cl}$.}
$\Omega$, which is obtained by
integrating $S_{cl}$ only over the world volume of one lattice cell and dividing
by its volume.
Having explained how to calculate the energy of a field configuration, in the next section we turn to the
problem of solving the equations of motion to find the ground state.

\section{Solving the equations}\label{sec:solequ}
\subsection{The comparison with Ginzburg-Landau theory}
Before turning to the equations of motion, it helps to first look at the
Ginzburg-Landau equations for an analogy.  In some suitable units
defined in~\cite{Abrikosov:1956sx,AbrikosovBook}, they are
\begin{align}
 & \left(-i \nabla - \vec{A}\right)^2 \psi -\psi
  + |\psi|^2 \psi = 0~,
  \label{eq:GLpsi} \\
  &~\nabla \times \nabla \times \vec{A} =
  -i\left(\bar{\psi} \nabla \psi - \psi \nabla \bar{\psi}\right)
  - |\psi|^2 \vec{A}~.
  \label{eq:GLA}
\end{align}
Only the structure of these
equations is important, so we have ignored constant factors.
Here $\psi$ is the wave function of Cooper pairs, and $\vec{A}$ is the
electromagnetic vector potential. The nine equations of motion in our system
can be split into two groups that roughly correspond to the two equations above.

The first of the two groups, hereafter called the condensate equations, contains
the six equations for the fields $\A^{1,2}_{x,y,u}$.
The superconducting condensate of
the dual field theory, which is like $\psi$ above,
is found by differentiating the on-shell action with respect to the boundary
values of $\A^{1,2}_{x,y}$, as in equation~\ref{eq:vevJmu}.
Of the six equations in this group, the
dynamical equations are for $\A^{1,2}_{x,y}$ and the
constraint\footnote{Recall that we have set $\A^a_u=0$. However, its
equations of motion still impose constraints on the other fields.
} equations are for $\A^{1,2}_u$.
So this first group is analogous to equation~\ref{eq:GLpsi}.
The analogy
can be made more clear. As mentioned above, we can make the
field definitions $\mathcal{E}_{x,y}=\A^1_{x,y}+i\A^2_{x,y}$. Doing so allows
us to combine the six real equations into three complex
equations, two dynamical and one constraint. The constraint equation relates
$\mathcal{E}_x$ and $\mathcal{E}_y$ such that there is only one complex degree
of freedom left, which is analogous to the state $\psi$. All this is hard to
see at the non-perturbative level, but it illustrates the strategy we follow for
solving the equations at each order: we use the constraint equation to reduce
the two dynamical equations into one, and then solve it.

The second group of equations, which we call the magnetic field equations, is for the fields $\A^3_{x,y,u}$, corresponding
to $\vec{A}$ in equation~\ref{eq:GLA} above. There are three such equations, one
of which is a constraint. At each order we will be able to use the constraint to
separate the equations into one for $\A^3_x$ and one for $\A^3_y$.

\subsection{The gauge field perturbative expansion in more detail}
Having defined the ansatz for our gauge potential in equation~\ref{eq:eqansatz}
we can learn more about the perturbative expansion by studying the non-linear
structure of the equations of motion. The equation for $\A^3_u$ is
\begin{align}
-\A^2_x \partial_u \A^1_x-\A^2_y \partial_u \A^1_y
+\A^1_x \partial_u \A^2_x+\A^1_y \partial_u \A^2_y
+\partial_y \partial_u \A^3_y+\partial_x\partial_u \A^3_x = 0~.
\end{align}
We see that the magnetic field components appear in the linear terms, while the
condensate components appear in quadratic terms. This suggests that a
contribution to the condensate components that is first order in the
perturbative expansion influences a second order contribution in the magnetic
field components. More generally, an odd order contribution to the condensate
components influences an even order contribution to the magnetic field
components.

This structure is common throughout all the equations of motion. It turns out
that terms in the perturbative expansion of the magnetic field components that
have an odd order vanish. The even order terms in the condensate components can
then also be set to zero. We can thus constrain the expansion ansatz of
equation~\ref{eq:eqansatz} to
\begin{align}
  \mathcal{E}_{x,y} &= \eps E_{x,y} + \eps^3 e_{x,y} + \bO(\eps^5)~,
  \nonumber \\
  \A^3_y &= x B_c + \eps^2 a^3_y + \bO(\eps^4)~,
  \label{eq:EAansatz}\\
  \A^3_x &= \eps^2 a^3_x + \bO(\eps^4)~.
  \nonumber
\end{align}
Here the calligraphic letters denote the non-perturbative fields. $E_{x,y}$ and
$e_{x,y}$ are first and third order contributions to the condensate components,
respectively, while $a^3_{x,y}$ are second order corrections to
$\A^3_{x,y}$.

Because of this convenient expansion of the fields, the condensate components
and the magnetic components decouple at each order. That means that at each
order, we only need to work with fields we have already solved at previous
orders. Our strategy is thus to solve for the fields in the following sequence:\\
\centerline{
\xymatrix{
  {\mathcal{E}_{x,y}} =~~~~~~~& *+[F]{\eps E_{x,y}} \ar[dr] &+&
  *+[F]{\eps^3 e_{x,y}} &+~~\bO(\eps^5)~,
  \\
  {
  \begin{array}{c}
    \A^3_y \\
    ~\\
    \A^3_x
  \end{array}
  }
  {
  \begin{array}{c}
    = \\
    ~\\
    =
  \end{array}
  }
  {
  \begin{array}{c}
    x B_c\\
    ~\\
    ~
  \end{array}
  }&
  {
  \begin{array}{c}
    +\\
    ~\\
    ~
  \end{array}
  }&
  *+[F]{
  \begin{array}{c}
    \eps^2 a^3_y\\
    ~\\
    \eps^2 a^3_x
  \end{array}
  } \ar@{->}[ur]&
  ~&
  {
  \begin{array}{c}
     +~~\bO(\eps^4)~,\\
    ~\\
     +~~\bO(\eps^4)~.
  \end{array}
  }
}
}\\
In the next section we start with the linear order solution, which will shed
more light on the procedure that must be implemented at higher orders.

\subsection{Solving the equations to linear order}
\label{sec:solvelinear}

Using the expansion~\ref{eq:EAansatz} and keeping terms to linear
order, we find that there are six remaining equations given (in complex form) by
\begin{align}
  0 &=
  -i B_c x \partial_u E_y - \partial_y \partial_u E_y
  - \partial_x \partial_u E_x~,
  \\
  0 &=
  B_c^2 x^2 E_x-i B_c
  E_y+\left(\frac{f}{u}-f'\right)
  \partial_u E_x-f \partial_u^2 E_x-2 i B_c x
  \partial_y E_x
  \nonumber \\
  &~
  -\partial_y^2 E_x+i B_c x
  \partial_x E_y+ \partial_x \partial_y E_y~,
  \\
  0 &=
  2 i B_c E_x+\left(\frac{f}{u}-f'\right)
  \partial_u E_y-f \partial_u^2 E_y+i B_c x
  \partial_x E_x
  +\partial_x \partial_y E_x - \partial_x^2 E_y~.
\end{align}
Here, as above, $f(u)=1-u^4$ for the AdS Schwarzschild model and $f(u)=1$ for
the hard wall model.

We can solve these equations by following Abrikosov \cite{AbrikosovBook}. The
solution is given by
\begin{align}
  E_y &= -i E_x~,
  \label{eq:solEy} \\
  E_x &= \sum_{n=-\infty}^\infty C_{n}
    e^{-inky-\frac{1}{2} B_c \left(x-\frac{n k}{B_c} \right)^2}
    U(u)~.
  \label{eq:solEx}
\end{align}
$U(u)$ is determined by solving
\begin{equation}
  U'' + \left(\frac{f'(u)}{f(u)}-\frac{1}{u} \right)U'
  + \frac{B_c}{f(u)} U
  =
  0~,
  \label{eq:Uu}
\end{equation}
subject to the constraints $U(0) = 0$ and $U'(1)=0$. For the AdS Schwarzschild
model ($f(u)=1-u^4$), the latter constraint comes from imposing regularity at
the horizon. It is possible to calculate
$B_c$ by numerically finding the value at which $U(u)$ satisfies these
constraints. There is an infinite tower of solutions to $B_c$, but we are only
interested in the lowest one, which is where the phase transition occurs.
For further details on solving this equation in the AdS Schwarzschild model,
see~\cite{Ammon:2011je}. For the hard wall model, $f(u)=1$ so the equation
simplifies to the extent that it can be solved analytically, the solution being
a Bessel function. Qualitatively the solutions for $U(u)$ in both models look
very similar and we are only interested in their numerical form. For the AdS
Schwarzschild model, we get
$B_c\approx 5.1$, while we get $B_c\approx 5.8$\footnote{This is the zero of the
Bessel function of the first kind $J_0(\sqrt{B})$.} for the hard wall model.

It should be noted that the solution~\ref{eq:solEx} for $E_x$ agrees precisely
(except for the factor of $U(u)$) with the linear order solution for the order
parameter close to the upper critical magnetic field $H_{c2}$ in the theory of
type II superconductors, as seen in~\cite{AbrikosovBook}. It is also the result
found by Chernodub et al. in~\cite{Chernodub:2011gs}.
Depending on the values of the parameters $C_n$ and $k$ (to be determined by the
higher order equations in the perturbative expansion), $E_x$ corresponds to
different inhomogeneous functions in
the $x,y$-plane. We are particularly interested in finding those with
lattice symmetries that represent evenly spaced vortices running in the $z$
direction in the gauge
theory.

\subsection{The Abrikosov lattice solution}\label{sec:abr}

Before going beyond linear order, we discuss the possible solutions we
can expect. The
number of coefficients specifying a configuration can make the problem of
finding the lowest energy solution unmanageable without making use of some
symmetries. We can argue that, since nothing in the setup is explicitly breaking
translational invariance in the $x,y$-directions, the solution should be a highly
symmetric lattice. A nice review of how lattices can be formed from the
Abrikosov solution (\ref{eq:solEx}) is given in~\cite{RosensteinLi2010}.
There the authors explain that in order for
$|E_x|$ to be a lattice solution, the coefficients $C_n$ must have the same
magnitude $|C_n|$ and moreover be periodic in some integer $P$, that is, $C_n =
C_{n+P}$.

In~\cite{Abrikosov:1956sx}, Abrikosov first studied the simplest solution, a
square lattice.  In this case, $P=1$, implying that $C_n = C$ for all $n$, and
$k=\sqrt{2\pi B_c}$. Later Kleiner et al. in~\cite{Kleiner:1964} generalised the
analysis by looking at $P=2$, with $C_1= \pm i C_0 = \pm i C$. This choice of
coefficients specifies a general rhombic lattice, with the shape of the rhombus
controlled by varying $k$. In particular, a square lattice can be
obtained by choosing $k=\sqrt{\pi B_c}$. This square lattice is the same as
Abrikosov's solution with $P=1$, but it is rotated by $\pi/4$ and translated. A
triangular lattice is obtained by choosing $k=3^{\frac{1}{4}}\sqrt{\pi B_c}$.

\begin{figure}
\centering
\includegraphics[width=0.4\textwidth]{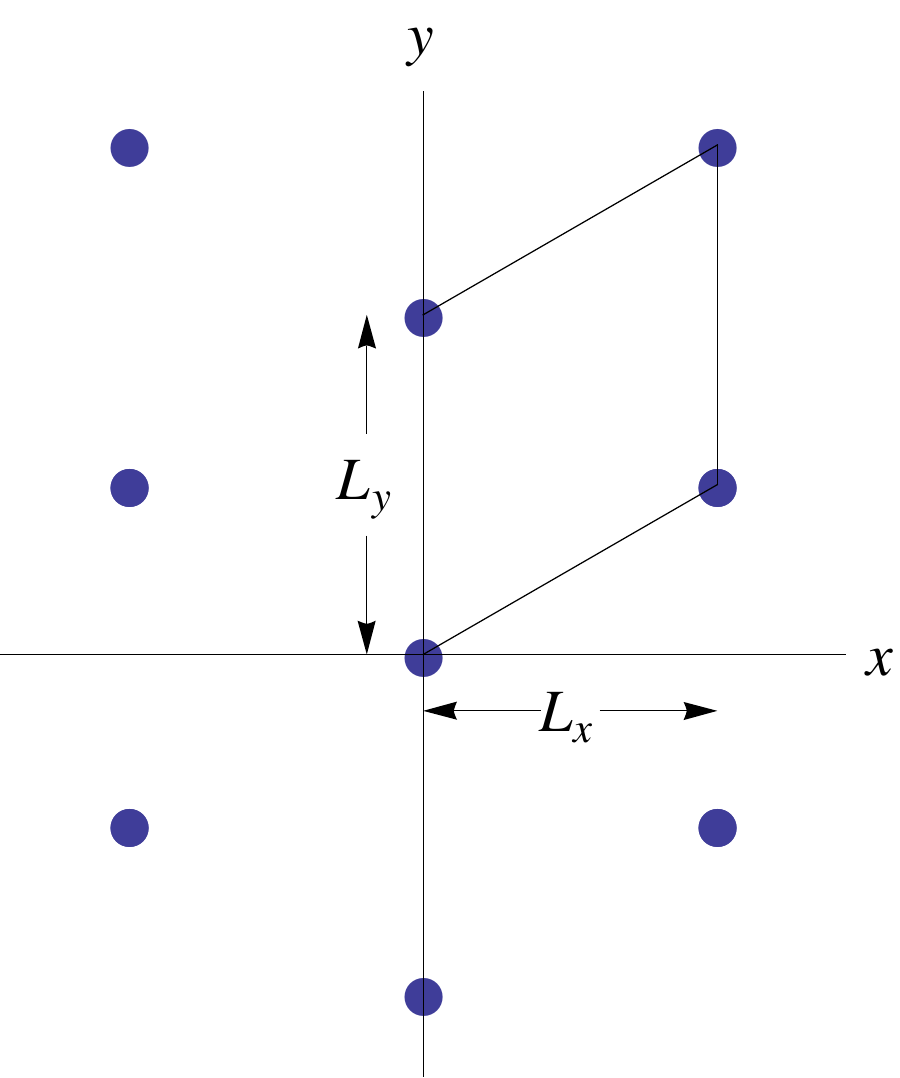}
\caption{A lattice cell, illustrating the meanings of $L_x$ and $L_y$ for a fixed area cell.
}
\label{fig:latticecell}
\end{figure}

To show how this works, we first substitute $P=2$ and $C_1=iC_0=iC$ into the
solution for $E_x$, which simplifies to
\begin{align}
  E_x &= C \sum_{n=-\infty}^\infty
    e^{i\frac{\pi}{2}n^2 -inky
      -\frac{1}{2} B_c \left(x-\frac{n k}{B_c} \right)^2}
    U(u)~.
  \label{eq:ExN2}
\end{align}
It is then easy to see the symmetries
$|E_x(x+[m+\frac{1}{2}q]L_x,y+[n+\frac{1}{2}q]L_y)|=|E_x(x,y)|$ for integers
$m$, $n$ and $q$. $L_x$ and $L_y$ are the lengths of the lattice cell in the
$x$ and $y$ directions, and are given by $L_x=2k/B_c$ and $L_y=2\pi/k$. See figure~\ref{fig:latticecell}.

We follow the approach of Kleiner et al, which is to compute the energy
density of the lattice for a range of values of the ratio
$L_x/L_y=k^2/\pi B_c$. This essentially means that we vary $k$.
The energy is computed numerically from the analytic expressions we obtain at
each order in the following sections.
What we find agrees with their result that the triangular lattice has the
lowest energy of the $P\le 2$ solutions.
When doing this, magnetic flux conservation is an important constraint. The total applied magnetic
field per unit area is constant, and each lattice cell corresponds to a vortex
with a single quantum of magnetic flux. This means that when comparing the
energy
of different lattices, we should make sure that they have the same
magnetic flux per unit area, which in turn means that their lattice cells have
the same area. Fortunately with this ansatz that is always the case since the
area $L_xL_y=4\pi/B_c$ is independent of $k$.

In the following sections we calculate analytic expressions for the higher
order corrections to the gauge field. We keep $P$ and the
coefficients $C_n$ general, except for imposing the periodicity
condition $C_n=C_{n+P}$.

\subsection{Higher order contributions to the energy}

In order to find the ground state solution we must calculate the energy of the
superconducting solutions and compare them to the normal phase
solution.  We can study the form of the energy as defined in section~\ref{sec:energy} to see how far we
must go in the perturbative expansion of the gauge fields.
The energy has terms that are quadratic and quartic in the gauge potential. The quartic
term ensures that the energy is bounded below, because it has a positive coefficient. The quartic terms have lowest
perturbative contributions of order $\eps^4$. One might expect
contributions of order $\eps^3$ coming from the zeroth order magnetic field
contribution multiplied by three first order corrections. However, from equation~\ref{eq:EAansatz} it can be shown that such terms do not arise. Thus we should expect to
expand to third order in $\A_{x,y}^{1,2}$ and fourth order in $\A_{x,y}^3$ .
However, it turns out that going to fourth order is not necessary because
inserting ansatz~\ref{eq:EAansatz} into the action of
equation~\ref{eq:actionym}, we find that the only fourth
order terms from $\A_{x,y}^3$ that appear at the fourth order of the action are
proportional to
$\sim \partial_y a^{(4)3}_x -\partial_x a^{(4)3}_y$. Here $a^{(4)3}_x$ and $ a^{(4)3}_y$ are the fourth order corrections to $\A^3_x$ and $\A^3_y$, respectively. This term respects the
lattice symmetries, thus on performing the integration over the lattice cell to get the free
energy density, it vanishes by Stokes' theorem.

We saw above that the parameters $k$ and $C_n$ in the solution~\ref{eq:solEx}
are not fixed by the equations of motion to linear
order. This is due to the fact that to linear order, the different vortices do not
interact. We can therefore not expect to fix any of the coefficients $C_n$ or
the spacing parameter $k$ at this order.
In fact, trying to calculate $\Delta \Omega$ to this order, which has no
quartic terms in $\A$, one finds that the free energy density is not
bounded below; increasing the overall magnitude of the condensate always
decreases $\Delta \Omega$.
To see which configuration, that is,
which set of values for $C_n$ and $k$, is energetically favourable, we clearly have to go beyond
linear order.
\subsection{Solving the equations to higher orders}
In this section we solve the equations of motion up to third order in the perturbation parameter.

The second order corrections to the gauge fields contribute to the potentials
$\A^3_x$ and $\A^3_y$, that is, $a^3_x$ and $a^3_y$ in~\ref{eq:EAansatz}. These fields source the external magnetic field and the magnetisation. We
impose that these corrections must vanish at the AdS boundary, so that the dual
field theory has a constant applied magnetic field. We find however that they do not
vanish throughout the bulk. In particular they develop non-vanishing
subleading terms in the boundary expansion, representing a magnetisation in the
field theory.

In appendix~\ref{app:eqa3xy} we explain how the equations for the Fourier modes of the
fields $a^3_x$ and $a^3_y$ can be decoupled. This yields the
following equations
\begin{align}
&u \partial_u\left(\frac{f}{u} \partial_u\hat{a}^3_{x,y}(m,n,u)\right)
-\left(k^2 n^2+\frac{4 B_c^2 m^2 \pi ^2}{k^2 P^2}\right) \hat{a}^3_{x,y}(m,n,u)\nonumber \\
&+T_{x,y}
 e^{-\frac{k^2 n^2}{4 B_c}+\frac{i n m \pi }{P}-\frac{B_c m^2 \pi
^2}{k^2 P^2}}  \left(\sum _{l=0}^{P-1} e^{\frac{2 i l m \pi }{P}}
\bar{C}_l C_{l+n}\right) U^2
= 0~,
\label{eq:tta3x}
\end{align}
where
\begin{equation}
T_x=-i\frac{\sqrt{B_c\pi}}{P} n~,~~~T_y=2i \frac{\pi ^{3/2}B_c^{3/2}}{k^2 P^2} m~,
\end{equation}
and
\begin{equation}
  a^3_{x,y}(x,y,u) = \sum_m \sum_n e^{-i\frac{2\pi m B_c}{Pk}x -inky}~
\hat{a}^3_{x,y}(m,n,u)~.
\end{equation}
As before, $P$ defines the periodicity in the $C_n$. The parameters $m$ and $n$
correspond to the Fourier space levels of these fields. In order to calculate the
solution $a_{x,y}^3(x,y,u)$ we will in theory need to solve these equations for
all values of $m$ and $n$. However, it will turn out to be sufficient to only
study the first few Fourier modes. The numerical procedure for solving these will
be explained in section~\ref{sec:numsol}

At third order we are studying the perturbative corrections to the condensate. Here we calculate
the corrections $e_x$ and $e_y$. It is reasonable to assume that the answer is
of the form
\begin{align}\label{eq:EC}
  \eps E_x + \eps^3 e_x &=
  \eps \sum_{n=-\infty}^\infty
  \left(C_{n} U(u) + \eps^2 c_{x,n}(u) \right)
    e^{-inky-\frac{1}{2} B_c \left(x-\frac{n k}{B_c} \right)^2}~,
  \\
  \eps E_y + \eps^3 e_y &=
  \eps \sum_{n=-\infty}^\infty
  \left(-iC_{n} U(u) + \eps^2 c_{y,n}(u) \right)
    e^{-inky-\frac{1}{2} B_c \left(x-\frac{n k}{B_c} \right)^2}~,
\end{align}
where we have made use of equation~\ref{eq:solEy} to relate the first order terms
$C_nU(u)$ in $\mathcal{E}_x$ and $\mathcal{E}_y$. $c_{(x,y),n}(u)$ is the first
perturbative correction to the condensate where the $u$ dependence is a function
of $n$ in contrast to the first order term.

We can write $e_x$ and $e_y$ in Fourier space, then
use the three condensate equations discussed in section~\ref{sec:solequ} to
calculate these corrections.
The one constraint equation can be used to decouple the other two equations. We
then have one equation for $c_{x,n}(u)$ and one for $c_{y,n}(u)$. Further
details are provided in appendix~\ref{app:eqcxyn}.

\subsection{Numerical solutions}\label{sec:numsol}

Having separated the equations into ordinary differential equations in $u$ by the
method outlined in the appendices, we can now solve them numerically. Both the
second
and third order equations take the same general form, given by
\begin{align}
  u \partial_u\left(\frac{f}{u} \partial_u\phi\right)
  + G(m,n) \phi + H(m,n,u) = 0~.
  \label{eq:ueqform}
\end{align}
This equation can be solved numerically by picking some parameters for $C_n$ and
$k$ that give a particular lattice and then using a shooting method to integrate
from
$u = 1$ (the horizon/hard wall cutoff) to $u = 0$ (the AdS boundary). It is an
inhomogeneous second order differential equation, so there are two integration
constants. The first is fixed by imposing regularity at the horizon or Neumann
boundary conditions at the hard wall cutoff. This fixes the value of $\partial_u
\phi(1)$. The second constant is obtained by demanding that $\phi(0)=0$, so that
the fields vanish at the AdS boundary. This vanishing corresponds to both the
magnetic field strength corrections and the source for the condensate being set to
zero. We fulfil this boundary condition by adjusting $\phi(1)$. Unlike in the case
of the first order equations, the equations here are not homogeneous and thus the
source sets a scale with which the value $\phi(1)$ can be compared. Changing
$\phi(1)$ in this case thus acts as more than just a scaling for the solution
and so is used as the tuning parameter to satisfy the UV constraint.

For all of the equations, we can implement this procedure for arbitrary integers
$m$ and $n$, corresponding to the different Fourier modes of the gauge fields.
This will then give a Fourier coefficient $\hat{a}^3_{x,y}(m,n,u)$ that can be
used to determine $a^3_{x,y}(x,y,u)$. Fortunately we do not have to do the
calculation for many different values of $m$ and $n$, because as the values get
large, the source term gets suppressed exponentially. This can be seen in
equation~\ref{eq:tta3x} for the second order terms and is true also for the third
order equation. For a vanishing source, the equations for $\hat{a}^3_x$ or
$\hat{a}^3_y$ have only the trivial solution. This means that
$\hat{a}^3_{x,y}(m,n,u)$ is negligibly small for large $m$ or $n$, and we can
therefore truncate the Fourier series for $a^3_{x,y}$ beyond $m,n\approx 3$.

\section{Results}\label{sec:result}

\subsection{Finding the minimum energy state}

\begin{figure}[!h]
\centering
\includegraphics[width=0.6\textwidth]{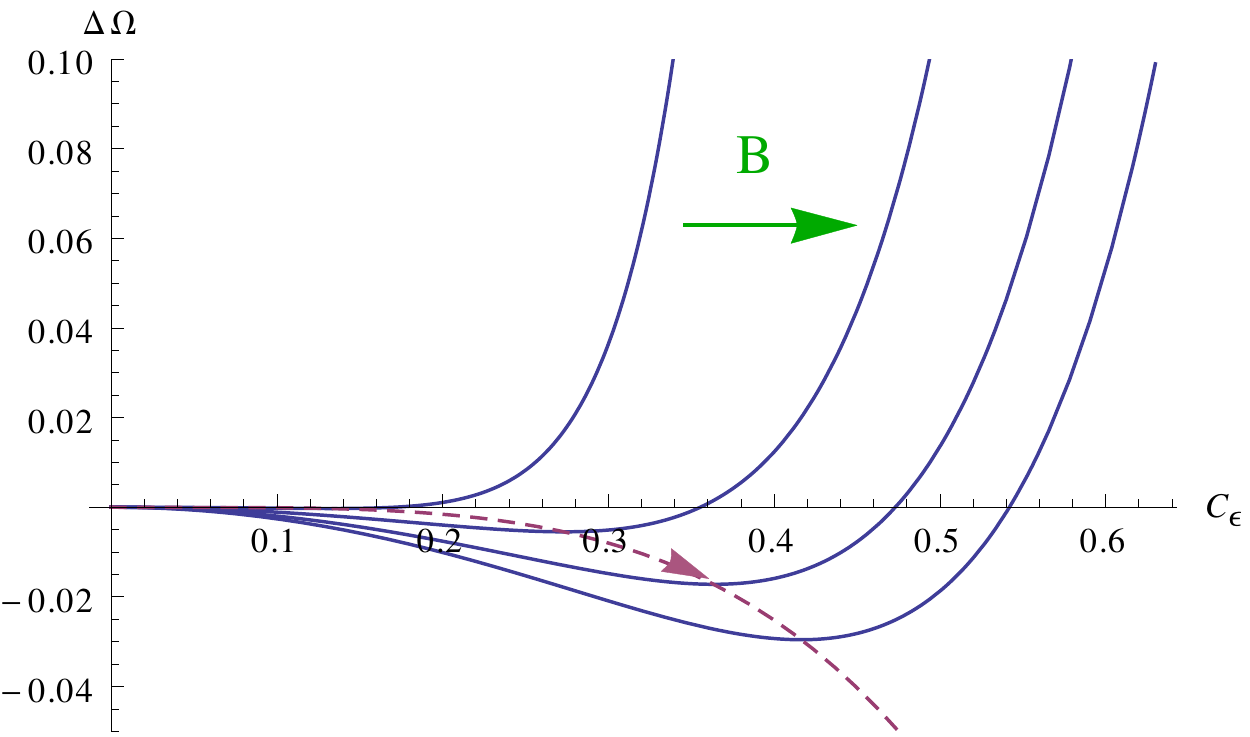}
\caption{The change in energy density in units of temperature as a function of
$C_\eps$, the overall condensate
scale. The leftmost curve corresponds to $B=B_c$, which is never negative for
nonzero condensate. Curves for $B<B_c$ are similar. Increasing $B$ beyond $B_c$
yields the curves to the right, and we see the formation of a clear minimum of
the energy
that is lower than the energy of the normal phase. The dashed line
traces out the minimum of each of these curves, which corresponds to the
energetically preferred size of the condensate as a function of $B$. This plot
was generated in the AdS Schwarzschild model for $P=2$ and
$k=3^{\frac{1}{4}}\sqrt{\pi B}$, corresponding to a triangular lattice. $B$
takes the values
$B\approx B_c, 1.04 B_c, 1.07 B_c, 1.1 B_c$ from left to right.
Changing $P$ and $k$ to correspond to different lattices or using the hard
wall model yields qualitatively similar results.
}
\label{fig:fec}
\end{figure}

\begin{figure}
\centering
\includegraphics[width=0.6\textwidth]{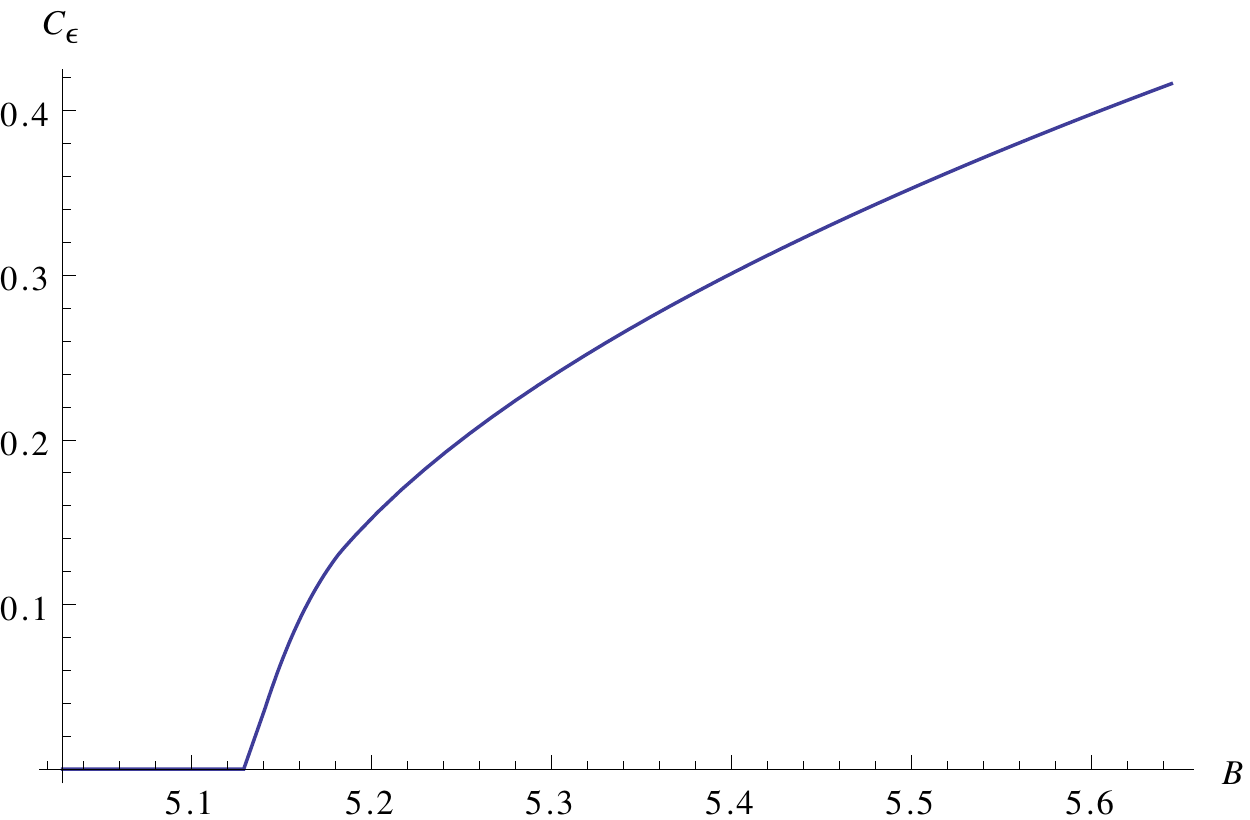}
\caption{$C_\eps\sim$ the overall condensate size for the AdS Schwarzschild solution in units of the temperature, as a function of the external
magnetic field $B$. For $B<B_c$, the condensate is zero, and for $B$ slightly
above $B_c$, we see a $(B-B_c)^{\frac{1}{2}}$ scaling behaviour.
This plot
was generated for $P=2$ and
$k=3^{\frac{1}{4}}\sqrt{\pi B}$, corresponding to a triangular lattice.
The plot for different lattices in both the AdS Schwarzschild and
hard wall models is the same, up to a scaling of the $B$ and $C_\eps$ axes.
For the triangular lattice, the AdS Schwarzschild model has scaling
behaviour $C_\eps = 0.58 (B-5.1)^{\frac{1}{2}}$
and the hard wall model has $C_\eps = 0.53 (B-5.8)^{\frac{1}{2}}$.
}
\label{fig:condensateB}
\end{figure}

As explained above, we wish to find the values of the parameters $k$, $P$ and
$C_n=C_{n+P}$ that give the minimum energy state. These parameters define the
shape of the lattice. Our analysis is only valid for $B$ slightly above $B_c$,
where $B_c$ was determined in section~\ref{sec:solvelinear}. The first step is
thus to pick a value for $B$ in this vicinity. We then choose a set of lattice
parameters that give us the lattice solution we wish to consider. As mentioned
in~\cite{RosensteinLi2010}, for lattice solutions all the $C_n$ must have the same
magnitude $C$. We can therefore fix $C_n$ up to the normalisation $C$, along with
a value of $k$, according to the discussion in section~\ref{sec:abr}. We then
substitute these values into the energy density that was defined in section~\ref{sec:energy}.
It takes the form
$\Delta\Omega=a_1\eps C+a_2\eps^2C^2+\dots~$. At this point we see that we can
redefine $C$ by absorbing a factor of $\eps$, which we call $C_\eps$.
$C_\eps$ is the only parameter left unfixed up to this point in the analysis. Here the $a_i$ are
values that are calculated numerically from substituting the solutions to the
equations of motion into the expression for the energy derived in
appendix~\ref{app:freeenergy}. $\Delta\Omega$ forms a
Mexican hat potential, which is easy to minimise numerically. An illustration
of this procedure is shown in figure~\ref{fig:fec}.
The plot in figure~\ref{fig:condensateB} shows the energy-minimising value of
$C_\eps$ as a function of magnetic
field near the phase transition at $B_c$. It shows that
$C_\eps \sim (B-B_c)^{\frac{1}{2}}$, so the condensate\footnote{Note that only
the combination $\eps C$ is physically relevant, not $C$ or $\eps$
independently.} has a critical exponent of $1/2$. A fit to the numerical data
for the triangular lattice
gives that $C_\eps = 0.58 (B-B_c)^{\frac{1}{2}}$ in the AdS Schwarzschild
model and $C_\eps = 0.53 (B-B_c)^{\frac{1}{2}}$ in the hard wall model.

\begin{figure}[!h]
\centering
\includegraphics[width=0.6\textwidth]{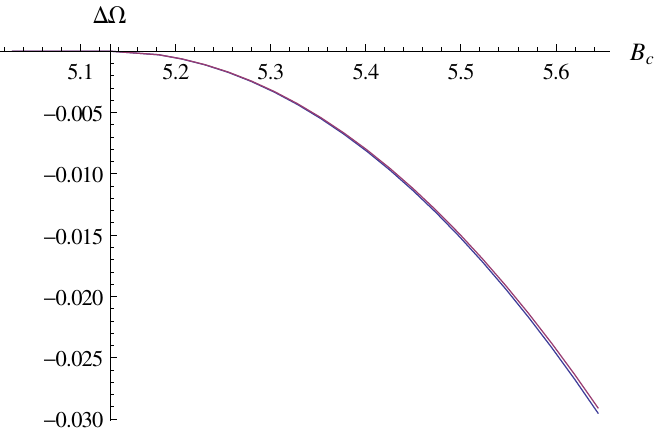}
\caption{The change in energy density (compared to the normal phase) for
the triangular and square lattices as the external applied magnetic field is
varied. The phase transition happens at $B_c\approx 5.1$, which is where the
coordinate axes are centred.
$\Delta \Omega_{\text{square}} - \Delta\Omega_{\text{triangle}}$ is so small
that the two plots are almost on top of each other. This is for the
AdS Schwarzschild model, but the plots for the hard wall model are identical
except for the scale on the axes. In the hard wall model, $B_c\approx 5.8$.
}
\label{fig:febsch}
\end{figure}

Having minimised with respect to $C_\eps$ for a given value of $B$ and a given lattice
configuration, we can plot the difference in the energy between the normal and
superconducting states. Figure~\ref{fig:febsch} shows $\Delta\Omega$, the
difference between the energy density in the superconducting and normal
phases, as a function of external magnetic field for two different lattices. The
first lattice is square, and the second is triangular. Both are described in
section~\ref{sec:abr}.

The curves in figure~\ref{fig:febsch} are the result of calculations in the
AdS Schwarzschild model, but we get the same results up to a rescaling of
the axes for the hard wall model. In the AdS Schwarzschild model, the critical
magnetic field $B_c\approx 5.1$, while in the hard wall model $B_c\approx 5.8$.
Each curve shows that the free energy density is
proportional to $\left(B-B_c\right)^2$. This shows that the phase transition is
second order, as expected if one looks
at the analogous case in Ginzburg-Landau theory. There one can
show (\cite{TinkhamBook}) that the free energy is proportional to
$\left(T-T_c\right)^2$, where $T_c$ is the phase transition critical
temperature.

\subsection{An analysis of $P=2$ solutions}
We now specialise to the case where the periodicity of the $C_n$ is $P=2$. This
describes a general rhombic lattice solution which includes both the triangular
and square lattices. The $P=1$ square lattice can be found within the $P=2$
solutions up to translation and rotation.  We here perform the analysis done
in~\cite{Kleiner:1964} as described at the end of section~\ref{sec:abr}.

\begin{figure}[!h]
\centering
\includegraphics[width=0.6\textwidth]{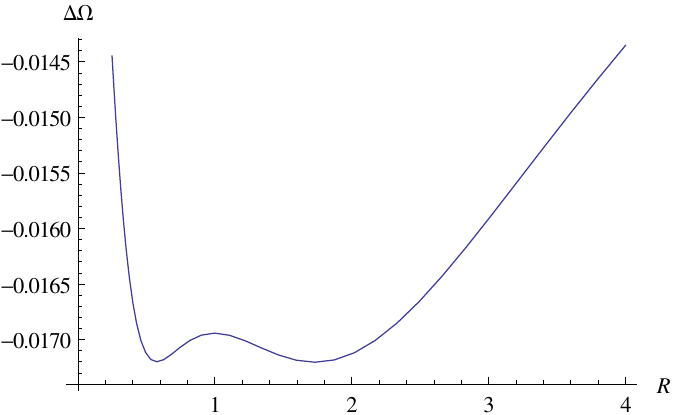}
\caption{The change in free
energy density as a function of $R=L_x/L_y$, the
ratio of side lengths of a constant area lattice cell. This plot is for the AdS
Schwarzschild model, but the plot for the hard wall model is identical up to a
rescaling of the axes. When
$R = 1$, the lattice is square
and the free energy achieves a local maximum. When $R=\sqrt{3}$ and $1/{\sqrt{3}}$, the lattice is
triangular and the free energy is at a global minimum. Note that the plot has
the symmetry $\Delta \Omega(R) = \Delta\Omega(1/R)$, which simply corresponds
to swapping the $x,y$-axes.
}
\label{fig:fersch}
\end{figure}

The energy difference as a function of $R$ is plotted in
figure~\ref{fig:fersch}.
By looking at the form of equation~\ref{eq:ExN2}, it is possible to
see that the triangular lattice occurs for $R=L_x/L_y=\sqrt{3}$ and
$R=1/\sqrt{3}$. In general, $R$ and $1/R$ give the same lattice but with the $x$
and $y$ directions flipped. This is why figure~\ref{fig:fersch} displays the
symmetry
$\Delta \Omega(R) =\Delta\Omega(1/R)$. The triangular lattice corresponds to a
global minimum of the energy as a function of $R$, as seen from the
figure. There is a local maximum for the square lattice, which is when $R=1$. As
$R\rightarrow\infty$ (or $R\rightarrow 0$), the free energy increases.
Intuitively one can understand this by making use of the properties of Abrikosov
vortices that we understand from type II superconductors. These vortices repel.
Since $R\rightarrow\infty$ and $R\rightarrow 0$ correspond to elongating the rhombic lattice cell
(while keeping the area constant) neighbouring vortices are squeezed together,
and since they repel, this is energetically unfavourable.

\begin{figure}[!h]
\centering
\includegraphics[width=0.6\textwidth]{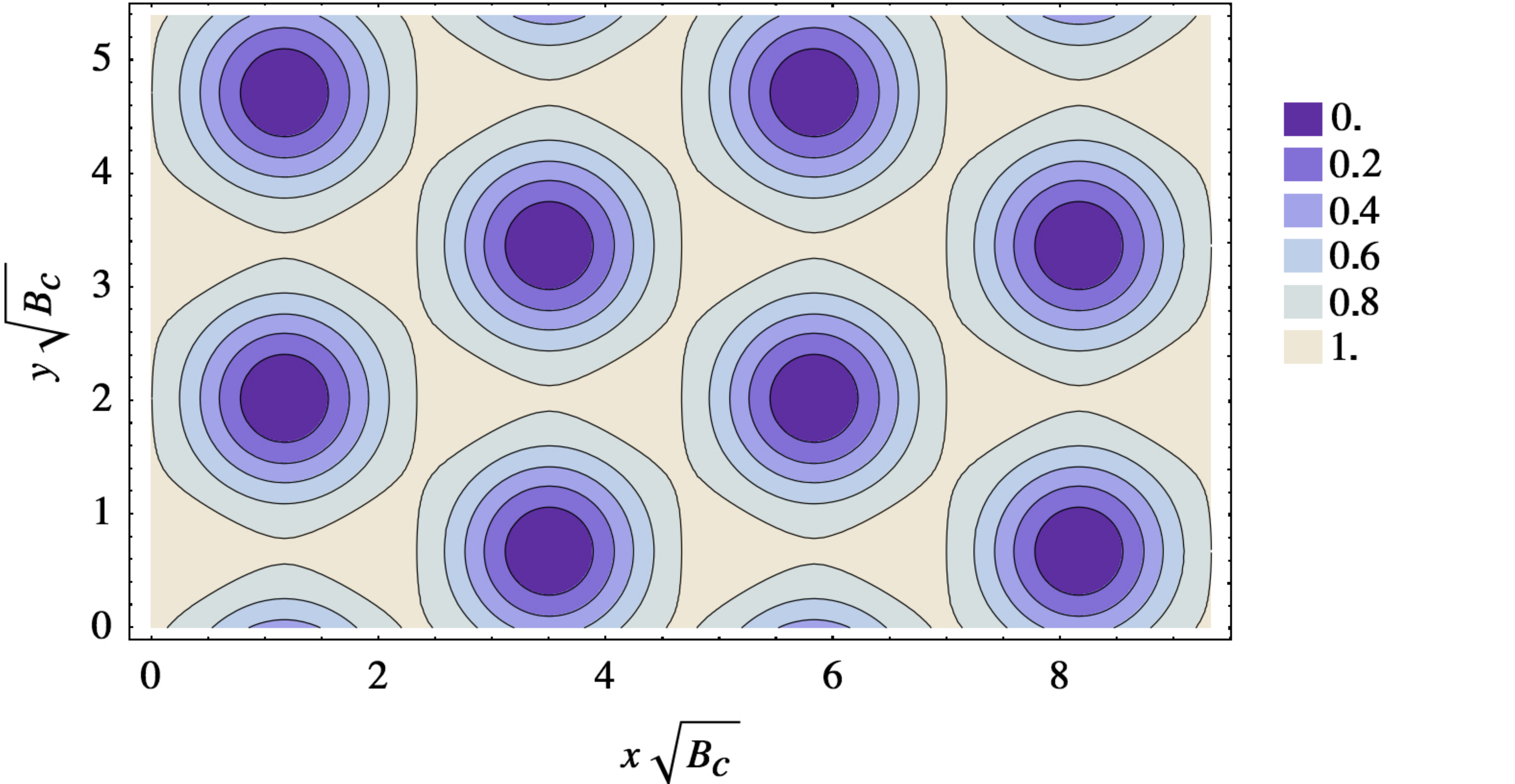}
\caption{A contour plot of
$3^{\frac{1}{4}} \sqrt{8}
\frac{\hat{g}^4}{L^2 U_\mathrm{sub}^2 C_\eps^2}
\left|\langle J^+_x \rangle\right|^2$,
the modulus squared of the field theory condensate dual to $E_x$ in the ground
state triangular
lattice. At the center of the dark
vortices, the condensate vanishes.
}
\label{fig:trianglecontour}
\end{figure}

We can calculate the condensate in the minimum energy state using
equation~\ref{eq:vevJmu}. The result, to linear order in $\eps$, is
\begin{equation}
\langle J^+_x \rangle
\equiv
\frac{\delta S_\mathrm{on-shell}}{\delta E^{(0)}_x}
=
\frac{L}{2 \hat{g}^2}
U_\mathrm{sub}
C_\eps \sum_{n=-\infty}^\infty
    e^{-i\frac{\pi}{2}n^2 +inky
      -\frac{1}{2} B_c \left(x-\frac{n k}{B_c} \right)^2}
\end{equation}
The AdS radius can be related to field theory quantities through the relation
$L^4= 2 \lambda {\alpha^\prime}^2$, where $\lambda$ is the 't Hooft coupling
and $\alpha'$ the string tension. The factor
$U_\mathrm{sub}$ is equal to the subleading term in the
boundary expansion of $U(u)$. Using equation~\ref{eq:Uu} it is possible to show
that
\begin{equation}
  U_\mathrm{sub}=B_c\int_0^{u_H} \frac{U(u)}{u} du~,
\end{equation}
so it can be determined numerically.
In figure~\ref{fig:trianglecontour} we present the contour plot of
$3^{\frac{1}{4}} \sqrt{8}
\frac{\hat{g}^4}{L^2 U_\mathrm{sub}^2 C_\eps^2}
\left|\langle J^+_x \rangle\right|^2$,
the modulus
squared of the condensate in the $x,y$-plane for the minimum energy solution
corresponding to the triangular lattice. The factors are chosen so that
the maximum value is 1. Substituting in the numerical values, we find that the
maximum value the condensate takes is
$|\langle J^+_x \rangle|=1.0\frac{L}{\hat{g}^2}\left(B-B_c\right)^\frac{1}{2}$
for the AdS Schwarzschild model, where $B_c\approx 5.1$, and
$|\langle J^+_x \rangle|=1.3\frac{L}{\hat{g}^2}\left(B-B_c\right)^\frac{1}{2}$
for the hard wall model, where $B_c\approx 5.8$.

We could also plot the magnetisation of the ground state, which is found from the
normalisable term in the boundary value expansion of $\partial_x a^3_y -
\partial_y a^3_x$. However, it takes the same form as the condensate and the numerics
indicate that it differs only up to a scale.

\section{Conclusion}\label{sec:conclu}

In this work we have found a likely ground state for the
black hole Yang-Mills instability analysed in~\cite{Ammon:2011je}. The solution,
being of a lattice form, clearly has much potential for analysis in
condensed matter models, where the breaking of translational invariance has
already been shown to be very important in getting realistic phenomenology. As we have explained, it also has possible implications for heavy ion collider physics. There are
a number of interesting areas where we could apply similar techniques and
perform further calculations to elucidate the phenomenology of the ground state
found here.

It is certainly important to understand exactly how universal this result is. We
have seen that the difference between the normal phase and superconducting phase energy  density of this solution is, up to a
scale, independent of the
background geometry in the two models that we have studied here. It would be very
useful to understand precisely where this universality stems from and find how
much we can deform the gravity solutions until the Abrikosov lattice is no longer
the ground state.

In the present work we have analysed lattices with $P=1$ and $P=2$, corresponding
to square and rhombic forms, respectively. Going to $P=3$ requires a large
increase in computational power. While this would be an interesting further
calculation, the analogous cases of type II superconductivity and the model of Chernodub et
al point to the triangular lattice being the true ground state. We thus
expect higher $P$ lattices to be energetically disfavoured.

Having found this solution, there are some extensions that can be made.
It will now be possible to study time dependent fluctuations about the ground state. In order to do this we would have to introduce a second perturbative parameter in addition to the parameter $\eps$ used in the current work. This would be analogous to the parameter $\alpha'$ in a D-brane construction. This would allow us to study the transport properties of the
lattice ground state, by looking at current-current correlation
functions. If we wish to study the effect of the lattice on the shear viscosity
to entropy density ratio, we would have to introduce gravitational back reaction in our model.
Clearly this will be a much more involved calculation, with many
non-linear couplings, but if we want to study the theory in a more realistic
scenario, where the stress energy tensor also has a lattice structure, such a
calculation would clearly be important.

It is expected that if the QCD vacuum is unstable to $\rho$ meson condensation
in extremely off-centre heavy ion collisions, then the timescale of the
instability would not be enough to form a well-defined lattice. Abrikosov
vortices may form, but the magnetic field would likely drop below the critical value
before they had time to arrange themselves into a lattice. It would be very
interesting to perform a real-time calculation in order study the formation
of the vortices and their movements as the magnetic field increased and
decreased through the lifetime of a single off-centre collision.

\section*{Acknowledgements}
Y.Y.B. was supported by MPG-CAS Doctoral Promotion Programme. The work of J.S. was supported by the European Union through a Marie
Curie Fellowship.
The work of J.E. and M.S. was supported in part by the DFG cluster of excellence
`Origin and Structure of the Universe' (www.universe-cluster.de). We thank Maxim Chernodub for useful discussions.

\appendix
\section{Deriving the equations for $a^3_{x,y}$}
\label{app:eqa3xy}

We substitute the ansatz~\ref{eq:EAansatz} into the full equations of
motion and
neglect terms beyond quadratic order in $\eps$. To get rid of all
appearances of $E_y$, we use the relation that $E_y=-i E_x$ from~\ref{eq:solEy}.
Then we find that there are only three equations in which the fluctuations
$a^3_{x,y}$ appear. We focus on those three equations.

The simplest equation of the three is the constraint equation, which came from
the equation of motion for $A^3_u$. To quadratic order, this equation
is simply
\begin{equation}
  \partial_u \partial_x a^3_x + \partial_u \partial_y a^3_y = 0~.
\end{equation}
The first thing to do is integrate by $u$. This gives an integration constant,
but by the fact that both $a^3_x$ and $a^3_y$ must vanish at $u=0$, this
integration constant vanishes. So the even simpler constraint is
\begin{equation}
    \partial_x a^3_x + \partial_y a^3_y = 0~.
\end{equation}
This is all we need to decouple the other two equations in $a^3_x$ and $a^3_y$.
These equations now become
\begin{align}
0 =& \frac{3}{2} E_x \partial_y \bar{E}_x+\frac{3}{2} \bar{E}_x
\partial_y E_x-\frac{1}{2} i E_x \partial_x \bar{E}_x+\frac{1}{2} i
\bar{E}_x \partial_x E_x
\nonumber \\
&+ u \partial_u \left( \frac{f}{u} \partial_u a^3_x \right)
+\partial^2_y a^3_x+\partial^2_x a^3_x~, \\
0=& -B_c x \bar{E}_x E_x-\frac{1}{2} i E_x  \partial_y
\bar{E}_x+\frac{1}{2} i \bar{E}_x  \partial_y E_x
\nonumber \\
&-\frac{3}{2} E_x
\partial_x \bar{E}_x-\frac{3}{2} \bar{E}_x \partial_x E_x
+ u \partial_u \left( \frac{f}{u} \partial_u a^3_y \right)
+\partial^2_y a^3_y+\partial^2_x a^3_y~,
\end{align}
which are partial differential equations with sources that come from the linear
order solutions. These
two equations only differ
by their source terms, so we will focus on $a^3_x$. $a^3_y$ should be
similar. Using the expression~\ref{eq:solEx}, we can see that the source term
is periodic with $y\sim y+\frac{2\pi}{k}$. $a^3_x$ must have the same
periodicity, so we can write it as a Fourier series,
\begin{align}
  a^3_x(x,y,u)=\sum_{n=-\infty}^\infty e^{-inky} \tilde{a}^3_x(x,n,u)~.
\end{align}
The equation becomes
\begin{align}
 \sum_m
  -i e^{-\frac{1}{2} B_c \left(
  -\frac{k m}{B_c}+x\right)^2-\frac{1}{2} B_c \left(-\frac{k (n+m)}{B_c}+x
\right)^2} k n
\bar{C}_{m} C_{n+m} U^2&~
\nonumber \\
-k^2 n^2 \tilde{a}^3_x
+u \partial_u \left( \frac{f}{u} \partial_u\tilde{a}^3_x \right)
+\partial^2_x \tilde{a}^3_x &= 0~.
\end{align}
We notice that the source term in this equation is periodic in the
$x$-direction; $x\sim x+\frac{Pk}{B_c}$. This lets us expand $\tilde{a}^3_x$ as
a Fourier series in $x$ as well:
\begin{align}
  \tilde{a}^3_x = \sum_m e^{-i \frac{2\pi m B_c}{Pk} x}
    \hat{a}^3_x(m, n, u)~.
\end{align}
Writing the source term as a series lets us then obtain the
equation~\ref{eq:tta3x} for the coefficients $\hat{a}^3_x(m, n, u)$.

Calling the source term $S(x)$, the na\"ive way of finding its Fourier
coefficients is to use the formula
\begin{align}
  \tilde{S}_n = \frac{B_c}{Pk}
  \int_0^{\frac{Pk}{B_c}} e^{i \frac{2\pi n B_c}{Pk} x}
    S(x)~.
\end{align}
However, the source terms contains Gaussians, and those are much easier to
integrate when the domain of integration is the entire real line. So we do the
following trick. Doing a continuous Fourier transform on a periodic function
gives a sum of $\delta$-functions,
\begin{align}
  \int dx~e^{ipx} S(x)
  &=
  \int dx~e^{ipx} \sum_m e^{-i \frac{2\pi m B_c}{Pk} x} \tilde{S}_n
  \nonumber \\
  &=
  2\pi \sum_m \tilde{S}_n \delta\left(p - \frac{2\pi m B_c}{Pk}\right)~.
  \label{eq:tildeSperiodic}
\end{align}
The coefficients in front of the $\delta$-functions are what we are looking
for. We get
\begin{align}
  \int dx~e^{ipx} S(x)
  &=
  -\sqrt{\frac{\pi }{B_c}} \sum_{m,n}
i e^{-\frac{k^2 n^2}{4 B_c}+\frac{i k m
p}{B_c}+\frac{i k n p}{2 B_c}-\frac{p^2}{4 B_c}} k n  \bar{C}_{m} C_{m+n}
U^2~.
\end{align}
Using
\begin{align}
  \sum_{m=-\infty}^\infty f(m) = \sum_{m=-\infty}^\infty\sum_{l=0}^{P-1} f(Pm+l)
\end{align}
and then using the symmetry $C_{i+P}=C_i$, the only $m$-dependence remaining in
the sum comes from $e^{\frac{ikPmp}{B_c}}$.
Making use of the identity
\begin{equation}
  \sum_{m=-\infty}^\infty e^{imq}
  = 2\pi \sum_{m=-\infty}^\infty \delta(q-2\pi m)
\end{equation}
and $\delta(\alpha x) = \frac{\delta(x)}{|\alpha|}$ gives us the sum over
$\delta$-functions from~\ref{eq:tildeSperiodic}. Then we can simply read off
the coefficients $\tilde{S}_n$. This gives us the equation~\ref{eq:tta3x}.

\section{Deriving the equations for $c_{x,n}$, $c_{y,n}$}
\label{app:eqcxyn}

The third order equations of motion are
\begin{align}
  0&=i a^3_x \partial_u E_x +a^3_y \partial_u E_x
  -i E_x \partial_u a^3_x-E_x \partial_u a^3_y
+i B_c x\partial_u e_y+\partial_y\partial_u e_y+\partial_x\partial_u e_x~,
  \label{eq:o3constraint}\\
  0&=
-i B_c x a^3_x E_x -2 B_c x a^3_y E_x-\bar{E}_x E_x^2
-a^3_x \partial_y E_x+2 i a^3_y \partial_y E_x
-a^3_y \partial_x E_x
\nonumber\\
&~-2 E_x \partial_y a^3_x+i E_x \partial_y a^3_y+E_x \partial_x a^3_y
+i B_c e_y -i B_c x \partial_x e_y-\partial_x\partial_y e_y
\nonumber\\
&~-B_c^2 x^2 e_x+2 i B_c x \partial_y e_x+\partial^2_y e_x
+u \partial_u \left( \frac{f}{u} \partial_u e_x \right)
\label{eq:o3e3x}\\
  0&=
B_c x a^3_x E_x+i \bar{E}_x E_x^2 -i a^3_x \partial_y E_x
+2 a^3_x \partial_x E_x-i a^3_y \partial_x E_x
\nonumber\\
&~+i E_x \partial_y a^3_x
+E_x \partial_x a^3_x-2 i E_x \partial_x a^3_y
-2 i B_c e_x-i B_c x \partial_x e_x-\partial_x\partial_y e_x
+\partial^2_x e_y
+u \partial_u \left( \frac{f}{u} \partial_u e_y \right)~.
\label{eq:o3e3y}
\end{align}
The first of these is the constraint equation. We use it to relate $e_x$ and
$e_y$. In order to do this, we first simplify it by noticing that, since
\begin{align}
  e_y &=
  \sum_{n=-\infty}^\infty
  c_{y,n}(u)
    e^{-inky-\frac{1}{2} B_c \left(x-\frac{n k}{B_c} \right)^2}~,
  \label{eq:e3yform}
\end{align}
we have that
$i B_c x\partial_u e_y+\partial_y\partial_u e_y = -i \partial_x \partial_u
e_y$. We can then integrate the entire equation by $u$, imposing vanishing
boundary conditions at the AdS boundary. The constraint equation then
simplifies to
\begin{align}
  0
  &=
-2 i \frac{E_x}{U} J_x-2 \frac{E_x}{U} J_y
+i a^3_x E_x +a^3_y E_x +\partial_x e_x
-i \partial_x e_y~,
\end{align}
where
\begin{align}
  J_{x,y}(x,y,u)=\int_0^u U(\tilde{u})
    \partial_{\tilde{u}} a^3_{x,y}(x,y,\tilde{u})
  d\tilde{u}~.
\end{align}
This allows us to eliminate $e_x$ in
equation~\ref{eq:o3e3y} (after differentiating it by $x$). We write each
function as a Fourier series in $y$ and find an equation for the coefficients
$c_{y,n}$. At this point the equation still has an $x$ dependence, which can be
eliminated by multiplying the equation by $(nk-B_cx)$ to
make it an even function in $x$ and then integrating $\int_{-\infty}^{\infty}
dx$. In doing so we use the solution for $E_x$ and the form for
$e_y$ given by~\ref{eq:e3yform}, as well as the Fourier series representation
of the other functions. Once this is done, we are left with an equation for
$e_y$ in the form~\ref{eq:ueqform}.

The resulting equation for $c_{y,n}$ is
\begin{align}
0&=
\sum_{q,r=-
\infty}^\infty\left\{
e^{-\frac{2 \pi  q \left(i k^2 P (n-r)+B_c
\pi  q\right)}{k^2 P^2}}
\left[
\frac{ C_{n-r} \left(-2 \left(k^2 P r+2 i B_c \pi  q\right)
\hat{J}_{x,qr} \right)}{k P}
\right. \right.
\nonumber \\
~&
\left.
+\frac{ C_{n-r} \left(\left(2 i k^2 P r-4 B_c \pi  q\right)
\hat{J}_{y,qr}+\left(2 i B_c \pi  q
\hat{a}^3_x+\left(-i k^2 P r+4 B_c \pi q\right) \hat{a}^3_y\right) U\right)}{k
P}\right]
\nonumber \\
~&
\left.
-\frac{i e^{-\frac{k^2 \left(3 r^2-3 r q+q^2\right)}{3 B_c}} \left(3 B_c+2 k^2 q
(-2 r+q)\right) \bar{C}_{n+q} C_{n+r} C_{n-r+q} U^3}{3 \sqrt{3} B_c}
\right\}
\nonumber \\
~&
-B_c c_{y,n}
+u \partial_u \left( \frac{f}{u} \partial_u c_{y,n} \right)~,
\end{align}
where
\begin{align}
  \hat{J}_{i,qr}(u)
  &= \int_0^u U(\tilde{u})
  \partial_{\tilde{u}} \hat{a}^3_i(q,r,\tilde{u})
  d\tilde{u}~,
\end{align}
for $i=x,y$.

A similar procedure gives the constraint equation in terms of the coefficients,
\begin{align}
  0
  &=
  c_{x,n}(u)-i c_{y,n}(u)
  \nonumber \\
  &~~~+\frac{1}{P k B_c}
  \sum_{q,r=-\infty}^\infty \Biggl\{
  e^{-\frac{2 \pi  q \left(i k^2 P (n-r)+B_c \pi q\right)}
    {k^2P^2}}
  \left(-i k^2 P r+2 B_c \pi  q\right) C_{n-r}
  \nonumber \\
  &~~~~~~~~~~~~~~~~~~~\times
  \left(2 \hat{J}_{x,qr}
  -2 i \hat{J}_{y,qr}-(\hat{a}^3_{x,qr}-i\hat{a}^3_{y,qr}) U(u)
  \right)
  \Biggr\}~.
\end{align}
Once the coefficients $c_{y,n}$ are found, we use this to calculate
$c_{x,n}$.

\section{Calculating the energy}
\label{app:freeenergy}

The difference between the energy of the superconducting phase
and that of the normal phase is
\begin{align}
  \Delta\mathcal{F}
  &=
  \frac{1}{4\hat{g}^2}\int d^5x \sqrt{-g} \left(
  \left.
  F^a_{\mu\nu}F^{a\mu\nu}\right|_{superconducting}
  -
  \left.
  F^a_{\mu\nu}F^{a\mu\nu}\right|_{normal}
 \right)~.
\end{align}
Note that for the AdS Schwarzschild model we implicitly divided by the
temperature to make the energy dimensionless.
We calculate the energy density by averaging over the domain
$0\le y < \frac{2\pi}{k}$, $0\le x < \frac{Pk}{B_c}$, $0\le u \le 1$ and
$t,z\in \mathbb{R}$. Since the integrand is independent of $t$ and $z$, the
averaging amounts to simply dropping the integration over those variables.
In the following expression we use
\begin{align}
  \mathcal{E}_{x,y}
  = \A^1_{x,y}+i\A^2_{x,y}
  = \sum_n \C_{(x,y),n}(u)
    e^{-ikny - \frac{1}{2} B_c \left(x-\frac{nk}{B_c}\right)^2}~,
\end{align}
we write $\A^3_x=a^3_x$ and
$\A^3_y=xB+a^3_y$, and call the averaged energy $\Delta\Omega$. The
result is
\begin{align}\label{eq:freeenergy1}
  4\hat{g}^2\Delta\Omega &=
  \int du \left\{
  \Omega_1(u) +
  \sum_{m,n=-\infty}^{\infty} \left[ \Omega_2(m,n,u) + \Omega_3(m,n,u) +
    \Omega_4(m,n,u)
  \right]
  \right.
  \nonumber \\
  &~
  \left.
    \sum_{m,n,p,q=-\infty}^{\infty} \Omega_5(m,n,q,r,u)
  \right\}~,
\end{align}
where
\begin{align}
  \Omega_1=&
  \frac{\sqrt{\pi B}}{k P u} \sum _{l=0}^{P-1}
  \frac{B}{2} \left(\sum_{j=x,y}
  \left(
    f \partial_u\bar{\C}_{j,l} \partial_u \C_{j,l}
      +\bar{\C}_{j,l} \C_{j,l}\right)+3 (i\bar{\C}_{y,l}
\C_{x,l}-i\bar{\C}_{x,l}  \C_{y,l})
  \right)~,
  \\
  \Omega_2=&
    \frac{1}{u}
    \left\|kn\hat{a}^3_x(m,n,u)-\frac{2Bm\pi}{kP}\hat{a}^3_y(m,n,u)\right\|^2
    +\frac{f}{u} \sum_{j=x,y} \left\|\partial_u \hat{a}^3_j(m,n,u)\right\|^2~,
  \\
\Omega_3=&
  \frac{\sqrt{\pi  B}}{2 k^2 P^2 u}\sum _{l=0}^{P-1}
    e^{-\frac{k^2 m^2}{4B}-\frac{i (2 l+m) n \pi }{P}
    -\frac{B n^2 \pi ^2}{k^2P^2}}
  \Bigl(
    \left(3 k^2 m P+2 i B n \pi \right) \hat{a}^3_x(n,m,u)
    \bar{\C}_{x,l+m}\C_{y,l}
  \nonumber \\
  &+\hat{a}^3_x(n,-m,u) \bar{\C}_{y,l} \left(\left(3 k^2 m P+2 i B n
\pi \right) \C_{x,l+m}-2 i k^2 m P \C_{y,l+m}\right)
  \nonumber \\
  &+\hat{a}^3_y(n,-m,u) \C_{x,l+m} \left(-4 i B n \pi  \bar{\C}_{x,l}
  +\left(i k^2 m P+6Bn \pi \right) \bar{\C}_{y,l}\right)
  \nonumber \\
  &+\hat{a}^3_y(n,m,u) \bar{\C}_{x,l+m} \C_{y,l}
  \left(-i k^2 m P-6 B n\pi\right)
  \Bigr)~,
  \\
  \Omega_4=&
  -\frac{1}{4 k P u}\sqrt{\frac{\pi B}{2}}
  e^{-\frac{k^2 \left(m^2+n^2\right)}{2B}}
  \times
  \nonumber \\
  &\sum _{l=0}^{P-1}
  \left(
    \bar{\C}_{y,l+m} \bar{\C}_{y,l+n} \C_{x,l} \C_{x,l+m+n}
      -2 \bar{\C}_{x,l+m} \bar{\C}_{y,l+n} \C_{x,l+m+n} \C_{y,l}
      +\bar{\C}_{x,l} \bar{\C}_{x,l+m+n} \C_{y,l+m} \C_{y,l+n}
  \right)~,
  \\
\Omega_5=&
  \frac{\sqrt{\pi B}}{P ku}
  \sum _{l=0}^{P-1}
    e^{-\frac{k^2 m^2}{4B}
      -\frac{i (2 l+m) n \pi }{P}
      -\frac{B n^2 \pi ^2}{k^2P^2}}
  \times
  \nonumber\\
  & \left(\hat{a}^3_y(n-q,-(m+r),u) \hat{a}^3_y(q,r,u)
    \bar{\C}_{x,l}\C_{x,l+m}
  -\hat{a}^3_x(n-q,r,u) \hat{a}^3_y(q,m-r,u)
    \bar{\C}_{x,l+m} \C_{y,l}
  \right.
  \nonumber\\
    &\left. -\hat{a}^3_x(n-q,r,u) \hat{a}^3_y(q,-(m+r),u)
    \bar{\C}_{y,l}\C_{x,l+m}
    +\hat{a}^3_x(n-q,-(m+r),u) \hat{a}^3_x(q,r,u)
    \bar{\C}_{y,l}\C_{y,l+m}\right)~.
\end{align}
In these expressions, $\C_{x,n}$ and $\C_{y,n}$ are functions of $u$.
Their complex conjugates are given by $\bar{\C}_{x,n}$ and $\bar{\C}_{y,n}$,
respectively. All the infinite sums in the energy~\ref{eq:freeenergy1} can be
terminated at a small finite value because of exponential suppression in the
$\Omega_{1\dots 5}$ terms.

In deriving this, it helps to make use of the formulae
\begin{align}
  \int_0^L dx &\sum_{m=-\infty}^\infty
  e^{-\frac{B_c}{2}\left(x-m L\right)^2}
  =
  \int_{-\infty}^\infty dx e^{-\frac{B_c}{2}x^2}~,
  \\
  \int_0^L dx &\sum_{m,n=-\infty}^\infty
  e^{-\frac{B_c}{2}\left(x-\frac{m L}{P}\right)^2
  -\frac{B_c}{2}\left(x-\frac{n L}{P}\right)^2}
  h(x,m,n)
  \\
  &=
  \int_{-\infty}^\infty dx \sum_{l=0}^{P-1} \sum_{m=-\infty}^\infty
  e^{-\frac{B_c}{2}\left(x-\frac{m L}{P}\right)^2
  -\frac{B_c}{2}\left(x-\frac{l L}{P}\right)^2}
  h(x,m,l)~,
\end{align}
where the latter is valid whenever $h(x,m,n)=h(x+L,m+P,n+P)$.

\bibliographystyle{JHEP}
\bibliography{references}

\providecommand{\href}[2]{#2}\begingroup\raggedright\begin{thebibliography}{10}

\bibitem{Ammon:2011je}
M.~Ammon, J.~Erdmenger, P.~Kerner, and M.~Strydom, {\it {Black Hole Instability
  Induced by a Magnetic Field}},  {\em Phys.Lett.} {\bf B706} (2011) 94--99,
  [\href{http://xxx.lanl.gov/abs/1106.4551}{{\tt arXiv:1106.4551}}].

\bibitem{Abrikosov:1956sx}
A.~Abrikosov, {\it {On the Magnetic properties of superconductors of the second
  group}},  {\em Sov.Phys.JETP} {\bf 5} (1957) 1174--1182.

\bibitem{Horowitz:2012ky}
G.~T. Horowitz, J.~E. Santos, and D.~Tong, {\it {Optical Conductivity with
  Holographic Lattices}},  {\em JHEP} {\bf 1207} (2012) 168,
  [\href{http://xxx.lanl.gov/abs/1204.0519}{{\tt arXiv:1204.0519}}].

\bibitem{Flauger:2010tv}
R.~Flauger, E.~Pajer, and S.~Papanikolaou, {\it {A Striped Holographic
  Superconductor}},  {\em Phys.Rev.} {\bf D83} (2011) 064009,
  [\href{http://xxx.lanl.gov/abs/1010.1775}{{\tt arXiv:1010.1775}}].

\bibitem{Chernodub:2010qx}
M.~Chernodub, {\it {Superconductivity of QCD vacuum in strong magnetic field}},
   {\em Phys.Rev.} {\bf D82} (2010) 085011,
  [\href{http://xxx.lanl.gov/abs/1008.1055}{{\tt arXiv:1008.1055}}].

\bibitem{Chernodub:2010zw}
M.~Chernodub, {\it {Electromagnetically superconducting phase of QCD vacuum
  induced by strong magnetic field}},  {\em AIP Conf.Proc.} {\bf 1343} (2011)
  149--151, [\href{http://xxx.lanl.gov/abs/1011.2658}{{\tt arXiv:1011.2658}}].

\bibitem{Chernodub:2011gs}
M.~Chernodub, J.~Van~Doorsselaere, and H.~Verschelde, {\it {Electromagnetically
  superconducting phase of vacuum in strong magnetic field: structure of
  superconductor and superfluid vortex lattices in the ground state}},
  \href{http://xxx.lanl.gov/abs/1111.4401}{{\tt arXiv:1111.4401}}.

\bibitem{Nielsen:1978rm}
N.~Nielsen and P.~Olesen, {\it {An Unstable Yang-Mills Field Mode}},  {\em
  Nucl.Phys.} {\bf B144} (1978) 376.

\bibitem{Ambjorn:1988tm}
J.~Ambjorn and P.~Olesen, {\it {On Electroweak Magnetism}},  {\em Nucl.Phys.}
  {\bf B315} (1989) 606.

\bibitem{Ambjorn:1989bd}
J.~Ambjorn and P.~Olesen, {\it {A Condensate Solution Of The Electroweak Theory
  Which Interpolates Between The Broken And The Symmetric Phase}},  {\em
  Nucl.Phys.} {\bf B330} (1990) 193.

\bibitem{Ambjorn:1989sz}
J.~Ambjorn and P.~Olesen, {\it {Electroweak Magnetism: Theory And
  Application}},  {\em Int.J.Mod.Phys.} {\bf A5} (1990) 4525--4558.

\bibitem{Domokos:2007kt}
S.~K. Domokos and J.~A. Harvey, {\it {Baryon number-induced Chern-Simons
  couplings of vector and axial-vector mesons in holographic QCD}},  {\em
  Phys.Rev.Lett.} {\bf 99} (2007) 141602,
  [\href{http://xxx.lanl.gov/abs/0704.1604}{{\tt arXiv:0704.1604}}].

\bibitem{Nakamura:2009tf}
S.~Nakamura, H.~Ooguri, and C.-S. Park, {\it {Gravity Dual of Spatially
  Modulated Phase}},  {\em Phys.Rev.} {\bf D81} (2010) 044018,
  [\href{http://xxx.lanl.gov/abs/0911.0679}{{\tt arXiv:0911.0679}}].

\bibitem{Chuang:2010ku}
W.-y. Chuang, S.-H. Dai, S.~Kawamoto, F.-L. Lin, and C.-P. Yeh, {\it {Dynamical
  Instability of Holographic QCD at Finite Density}},  {\em Phys.Rev.} {\bf
  D83} (2011) 106003, [\href{http://xxx.lanl.gov/abs/1004.0162}{{\tt
  arXiv:1004.0162}}].

\bibitem{Bergman:2011rf}
O.~Bergman, N.~Jokela, G.~Lifschytz, and M.~Lippert, {\it {Striped instability
  of a holographic Fermi-like liquid}},  {\em JHEP} {\bf 1110} (2011) 034,
  [\href{http://xxx.lanl.gov/abs/1106.3883}{{\tt arXiv:1106.3883}}].

\bibitem{Bayona:2011ab}
C.~B. Bayona, K.~Peeters, and M.~Zamaklar, {\it {A Non-homogeneous ground state
  of the low-temperature Sakai-Sugimoto model}},  {\em JHEP} {\bf 1106} (2011)
  092, [\href{http://xxx.lanl.gov/abs/1104.2291}{{\tt arXiv:1104.2291}}].

\bibitem{Takeuchi:2011uk}
S.~Takeuchi, {\it {Modulated Instability in Five-Dimensional U(1) Charged AdS
  Black Hole with R**2-term}},  {\em JHEP} {\bf 1201} (2012) 160,
  [\href{http://xxx.lanl.gov/abs/1108.2064}{{\tt arXiv:1108.2064}}].

\bibitem{Ooguri:2010kt}
H.~Ooguri and C.-S. Park, {\it {Holographic End-Point of Spatially Modulated
  Phase Transition}},  {\em Phys.Rev.} {\bf D82} (2010) 126001,
  [\href{http://xxx.lanl.gov/abs/1007.3737}{{\tt arXiv:1007.3737}}].

\bibitem{Donos:2012gg}
A.~Donos and J.~P. Gauntlett, {\it {Helical superconducting black holes}},
  {\em Phys.Rev.Lett.} {\bf 108} (2012) 211601,
  [\href{http://xxx.lanl.gov/abs/1203.0533}{{\tt arXiv:1203.0533}}].

\bibitem{Donos:2012wi}
A.~Donos and J.~P. Gauntlett, {\it {Black holes dual to helical current
  phases}},  {\em Phys.Rev.} {\bf D86} (2012) 064010,
  [\href{http://xxx.lanl.gov/abs/1204.1734}{{\tt arXiv:1204.1734}}].

\bibitem{Ammon:2011hz}
M.~Ammon, J.~Erdmenger, S.~Lin, S.~Muller, A.~O'Bannon, et~al., {\it {On
  Stability and Transport of Cold Holographic Matter}},  {\em JHEP} {\bf 1109}
  (2011) 030, [\href{http://xxx.lanl.gov/abs/1108.1798}{{\tt
  arXiv:1108.1798}}].

\bibitem{Donos:2011qt}
A.~Donos, J.~P. Gauntlett, and C.~Pantelidou, {\it {Spatially modulated
  instabilities of magnetic black branes}},  {\em JHEP} {\bf 1201} (2012) 061,
  [\href{http://xxx.lanl.gov/abs/1109.0471}{{\tt arXiv:1109.0471}}].

\bibitem{Donos:2011bh}
A.~Donos and J.~P. Gauntlett, {\it {Holographic striped phases}},  {\em JHEP}
  {\bf 1108} (2011) 140, [\href{http://xxx.lanl.gov/abs/1106.2004}{{\tt
  arXiv:1106.2004}}].

\bibitem{Bolognesi:2010nb}
S.~Bolognesi and D.~Tong, {\it {Monopoles and Holography}},  {\em JHEP} {\bf
  1101} (2011) 153, [\href{http://xxx.lanl.gov/abs/1010.4178}{{\tt
  arXiv:1010.4178}}].

\bibitem{Sutcliffe:2011sr}
P.~Sutcliffe, {\it {Monopoles in AdS}},  {\em JHEP} {\bf 1108} (2011) 032,
  [\href{http://xxx.lanl.gov/abs/1104.1888}{{\tt arXiv:1104.1888}}].

\bibitem{Allahbakhshi:2011nh}
D.~Allahbakhshi, {\it {On Holography of Julia-Zee Dyon}},  {\em JHEP} {\bf
  1109} (2011) 085, [\href{http://xxx.lanl.gov/abs/1105.3677}{{\tt
  arXiv:1105.3677}}].

\bibitem{Maeda:2009vf}
K.~Maeda, M.~Natsuume, and T.~Okamura, {\it {Vortex lattice for a holographic
  superconductor}},  {\em Phys.Rev.} {\bf D81} (2010) 026002,
  [\href{http://xxx.lanl.gov/abs/0910.4475}{{\tt arXiv:0910.4475}}].

\bibitem{Domenech:2010nf}
O.~Domenech, M.~Montull, A.~Pomarol, A.~Salvio, and P.~J. Silva, {\it {Emergent
  Gauge Fields in Holographic Superconductors}},  {\em JHEP} {\bf 1008} (2010)
  033, [\href{http://xxx.lanl.gov/abs/1005.1776}{{\tt arXiv:1005.1776}}].

\bibitem{Murray:2011gr}
J.~M. Murray and Z.~Tesanovic, {\it {Isolated Vortex and Vortex Lattice in a
  Holographic p-wave Superconductor}},  {\em Phys.Rev.} {\bf D83} (2011)
  126011, [\href{http://xxx.lanl.gov/abs/1103.3232}{{\tt arXiv:1103.3232}}].

\bibitem{Gubser:2008wv}
S.~S. Gubser and S.~S. Pufu, {\it {The Gravity dual of a p-wave
  superconductor}},  {\em JHEP} {\bf 0811} (2008) 033,
  [\href{http://xxx.lanl.gov/abs/0805.2960}{{\tt arXiv:0805.2960}}].

\bibitem{Ammon:2008fc}
M.~Ammon, J.~Erdmenger, M.~Kaminski, and P.~Kerner, {\it {Superconductivity
  from gauge/gravity duality with flavor}},  {\em Phys.Lett.} {\bf B680} (2009)
  516--520, [\href{http://xxx.lanl.gov/abs/0810.2316}{{\tt arXiv:0810.2316}}].

\bibitem{Ammon:2009fe}
M.~Ammon, J.~Erdmenger, M.~Kaminski, and P.~Kerner, {\it {Flavor
  Superconductivity from Gauge/Gravity Duality}},  {\em JHEP} {\bf 0910} (2009)
  067, [\href{http://xxx.lanl.gov/abs/0903.1864}{{\tt arXiv:0903.1864}}].

\bibitem{Chunlen:2012zy}
S.~Chunlen, K.~Peeters, P.~Vanichchapongjaroen, and M.~Zamaklar, {\it
  {Instability of N=2 gauge theory in compact space with an isospin chemical
  potential}},  {\em JHEP} {\bf 1301} (2013) 035,
  [\href{http://xxx.lanl.gov/abs/1210.6188}{{\tt arXiv:1210.6188}}].

\bibitem{Djukanovic:2005ag}
D.~Djukanovic, M.~R. Schindler, J.~Gegelia, and S.~Scherer, {\it {Quantum
  electrodynamics for vector mesons}},  {\em Phys.Rev.Lett.} {\bf 95} (2005)
  012001, [\href{http://xxx.lanl.gov/abs/hep-ph/0505180}{{\tt
  hep-ph/0505180}}].

\bibitem{Chernodub:2011mc}
M.~Chernodub, {\it {Spontaneous electromagnetic superconductivity of vacuum in
  strong magnetic field: evidence from the Nambu--Jona-Lasinio model}},  {\em
  Phys.Rev.Lett.} {\bf 106} (2011) 142003,
  [\href{http://xxx.lanl.gov/abs/1101.0117}{{\tt arXiv:1101.0117}}].

\bibitem{Skokov:2009qp}
V.~Skokov, A.~Y. Illarionov, and V.~Toneev, {\it {Estimate of the magnetic
  field strength in heavy-ion collisions}},  {\em Int.J.Mod.Phys.} {\bf A24}
  (2009) 5925--5932, [\href{http://xxx.lanl.gov/abs/0907.1396}{{\tt
  arXiv:0907.1396}}].

\bibitem{Bzdak:2011yy}
A.~Bzdak and V.~Skokov, {\it {Event-by-event fluctuations of magnetic and
  electric fields in heavy ion collisions}},  {\em Phys.Lett.} {\bf B710}
  (2012) 171--174, [\href{http://xxx.lanl.gov/abs/1111.1949}{{\tt
  arXiv:1111.1949}}].

\bibitem{Callebaut:2011ab}
N.~Callebaut, D.~Dudal, and H.~Verschelde, {\it {Holographic rho mesons in an
  external magnetic field}},  \href{http://xxx.lanl.gov/abs/1105.2217}{{\tt
  arXiv:1105.2217}}.

\bibitem{Witten:1998qj}
E.~Witten, {\it {Anti-de Sitter space and holography}},  {\em
  Adv.Theor.Math.Phys.} {\bf 2} (1998) 253--291,
  [\href{http://xxx.lanl.gov/abs/hep-th/9802150}{{\tt hep-th/9802150}}].

\bibitem{Erlich:2005qh}
J.~Erlich, E.~Katz, D.~T. Son, and M.~A. Stephanov, {\it {QCD and a holographic
  model of hadrons}},  {\em Phys.Rev.Lett.} {\bf 95} (2005) 261602,
  [\href{http://xxx.lanl.gov/abs/hep-ph/0501128}{{\tt hep-ph/0501128}}].

\bibitem{DaRold:2005zs}
L.~Da~Rold and A.~Pomarol, {\it {Chiral symmetry breaking from five dimensional
  spaces}},  {\em Nucl.Phys.} {\bf B721} (2005) 79--97,
  [\href{http://xxx.lanl.gov/abs/hep-ph/0501218}{{\tt hep-ph/0501218}}].

\bibitem{AbrikosovBook}
A.~Abrikosov, {\em Fundamentals of the Theory of Metals}.
\newblock North-Holland, Amsterdam, 1988.

\bibitem{RosensteinLi2010}
B.~Rosenstein and D.~Li, {\it Ginzburg-landau theory of type ii superconductors
  in magnetic field},  {\em Rev. Mod. Phys.} {\bf 82} (Jan, 2010) 109--168.

\bibitem{Kleiner:1964}
W.~H. Kleiner, L.~M. Roth, and S.~H. Autler, {\it Bulk solution of
  ginzburg-landau equations for type ii superconductors: Upper critical field
  region},  {\em Phys. Rev.} {\bf 133} (Mar, 1964) A1226--A1227.

\bibitem{TinkhamBook}
M.~Tinkham, {\em Introduction to Superconductivity}.
\newblock Robert E. Krieger Publishing Company, Malabar, Florida, 1980.

\end{thebibliography}\endgroup

\end{document}